\documentclass[aps,superscriptaddress,prl,twocolumn,amsmath, amssymb,
]
{revtex4-2}

\pdfoutput=1

\usepackage{graphicx}
\usepackage[mathlines]{lineno}
\usepackage[svgnames]{xcolor}
\usepackage{hyperref}
\usepackage{textcomp}
\hypersetup{
	colorlinks = true,
	linkcolor = Blue,
	citecolor = Blue,
	urlcolor  = Blue
}
\newcommand{\spa}{SrPtAs}
\newcommand{\zri}{Zr$_3$Ir}
\newcommand{\musr}{$\mu$SR}

\newcommand{\SRO}{Sr$_{2}$RuO$_{4}$}

\usepackage{comment}
\usepackage{ltablex}
\usepackage{multirow}
\usepackage{threeparttable}
\usepackage[referable]{threeparttablex}
\usepackage[caption=false]{subfig}
\usepackage{pbox}
\usepackage{booktabs,microtype,afterpage} 
\usepackage{ulem}
\begin{document}

\title{Intrinsic nature of spontaneous magnetic fields in superconductors with time-reversal symmetry breaking} 

\author{B. M. Huddart}
\email{benjamin.m.huddart@durham.ac.uk}
\affiliation{Centre for Materials Physics, Durham University, Durham DH1 3LE, United Kingdom}
\author{I. J. Onuorah}
\email{ifeanyijohn.onuorah@unipr.it}
\affiliation{Department of Mathematical, Physical and Computer Sciences, University of Parma, 43124 Parma, Italy}
\author{M. M. Isah}
\affiliation{Department of Mathematical, Physical and Computer Sciences, University of Parma,  43124 Parma, Italy}
\author{P. Bonf\`{a}}
\affiliation{Department of Mathematical, Physical and Computer Sciences, University of Parma,  43124 Parma, Italy}
\author{S. J. Blundell}
\affiliation{Department of Physics, Clarendon Laboratory, Oxford University, Parks Road, Oxford OX1 3PU, United Kingdom}
\author{S. J. Clark}
\affiliation{Centre for Materials Physics, Durham University, Durham DH1 3LE, United Kingdom}
\author{R. De Renzi}
\affiliation{Department of Mathematical, Physical and Computer Sciences, University of Parma,  43124 Parma, Italy}
\author{T. Lancaster}
\affiliation{Centre for Materials Physics, Durham University, Durham DH1 3LE, United Kingdom}

\date{\today}

\begin{abstract}
We present a systematic investigation of  muon-stopping states in superconductors that reportedly exhibit spontaneous magnetic fields below their transition temperatures due to time-reversal symmetry breaking.  These materials include elemental rhenium, several intermetallic systems and Sr$_2$RuO$_4$. We demonstrate that the presence of the muon leads to only a limited and relatively localized perturbation to the local crystal structure, while any small changes to the electronic structure occur several electron volts below the Fermi energy leading to only minimal changes in the charge density on ions close to the muon. Our results imply that the muon-induced perturbation alone is unlikely to lead to the observed spontaneous fields in these materials, whose origin is more likely intrinsic to the time-reversal symmetry broken superconducting state.
\end{abstract}

\maketitle 

A crucial issue in resolving the mechanism for unconventional superconductivity is the presence or absence of time-reversal symmetry breaking (TRSB) \cite{Ghosh_2020}, a property that can provide a tight constraint on the symmetry of the superconducting gap.  The conventional s-wave singlet BCS pairing conserves time-reversal symmetry, but triplet pairing does not.  For example, a p-wave gap symmetry was ascribed \cite{mackenzie-rmp,sigrist2005} to the superconductor Sr$_2$RuO$_4$ on the basis of its supposed triplet order parameter deduced from NMR \cite{ishida1998} and the presence of time-reversal symmetry breaking deduced from muon-spin relaxation ($\mu$SR) \cite{luke1998} and polar Kerr effect \cite{xia2006} measurements.  However, the triplet nature of Sr$_2$RuO$_4$ has recently been discounted on the basis of a new NMR investigation \cite{pustogow2019}, thereby reopening the question about the nature of the gap symmetry in this compound, with several alternative singlet gap structures proposed \cite{kivelson2020,romer2020,benhabib2021,Grinenko2021}, each of which would be consistent with TRSB. There is also a recent suggestion that the experimental signature of TRSB is not intrinsic but originates from  inhomogeneous strain fields near edge dislocations \cite{willa2020inhomogeneous}.  

The nature of TRSB superconductivity, and the need to understand how it is detected, is a question with a much wider applicability than merely the particular case of Sr$_2$RuO$_4$.  This is because the appearance of spontaneous magnetic fields is found in a large collection of superconductors using $\mu$SR measurements, though importantly it is absent for most superconductors (SCs) \cite{Ghosh_2020}. 
\musr\ has emerged as an effective probe of superconducting properties \cite{Sonier2000}, extracting the penetration depth and hence the superfluid stiffness \cite{Uemura1989}, examining vortex lattice melting \cite{Lee1993,Lee1997} and determining the nature of the pairing \cite{Sonier1994,biswas2018}.  The superconducting vortex lattice produced by an {\sl applied field} is, in general, incommensurate with the crystalline lattice and so the precise location of the muon site makes no difference in these studies.  Wherever the muon sits inside the unit cell, it will uniformly sample the magnetic field distribution produced by the vortex lattice \cite{Brandt1988}.  This is not the case for muon studies of magnetism for which the local field extracted by \musr\ depends sensitively on the location of the muon site (see e.g.\ \cite{Amato2014}).  This issue becomes extremely relevant for superconductors studied in {\sl zero applied magnetic field} where the signature of TRSB is the appearance of a very small spontaneous local field.  
There is currently no accepted theory which predicts how large the spontaneous field should be, to what extent these spontaneous fields should be screened by supercurrents, whether these fields are particularly associated with defects, interfaces and domain boundaries, or indeed whether the presence of the muon itself might play the role of a defect.  Nevertheless, the results of \musr\ experiments have been used to argue for TRSB on the basis of spontaneous fields detected in a number of unconventional superconductors, including \SRO \cite{luke1998}, LaNiC$_2$ \cite{PhysRevLett.102.117007}, SrPtAs \cite{biswas2013}, Zr$_3$Ir \cite{shang2019,sajilesh2019},  Re  \cite{PhysRevLett.121.257002}, and Re$_6$Zr \cite{PhysRevLett.112.107002}.
In this Letter, we critically reexamine these experiments by calculating the muon site in these candidate TRSB superconductors using density functional theory (DFT), to assess the degree to which the muon perturbs its local environment. 

\footnotetext{See Supplemental Material at [URL will be inserted by publisher] for further details of the the density functional theory calculations carried out on each of the materials in this study, which includes Refs. \cite{CASTEP,QE-2009,QE-2017,pbe1996,mpgrid1976,hamann2013,ultsft1990,gbrv2014,vesta2008,hao2014,oguchi1995,singh1995,schmidt1996,hideseek,marzari1999,wenski1986,PhysRevB.80.094524,De_Renzi_2012,Lamura_2013,lowdin1950,Cenzual1985,PhysRevB.94.214513,PhysRevB.27.6121,PhysRevB.37.4425,PhysRevB.43.3284,gygax}}
\begin{figure}[hb]
	\includegraphics[width=\columnwidth]{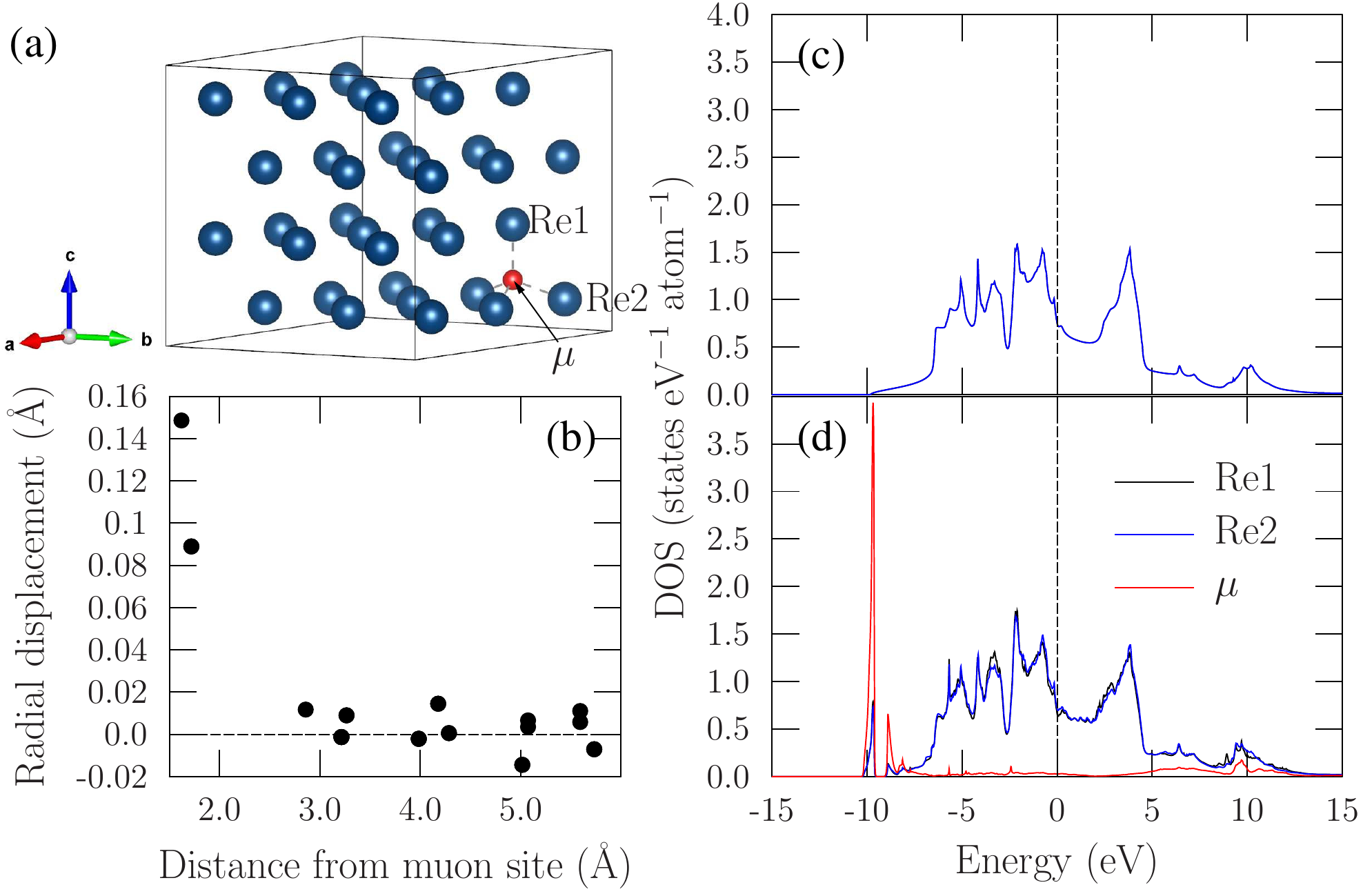}
	\caption{Tetrahedral muon site in  rhenium. (a) The two distinct nearest-neighbour Re environments, Re1 and Re2, in the coordination tetrahedron of the muon. (b) Radial displacements of the Re atoms as a function of their distances from the muon. PDOS for the nearest-neighbor Re atoms (c) without a muon and (d) with a muon. Energies are given with respect to the Fermi energy.}
	\label{fig:fig1}
\end{figure}

We first consider muon stopping sites in elemental rhenium, which is the material exhibiting spontaneous magnetic fields with the simplest crystal structure.
Structural relaxations of a supercell containing a muon yield two crystallographically distinct muon sites and, as an illustration of the physics relevant to the more complicated materials discussed below, 
we first consider the site tetrahedrally coordinated by Re atoms. For each of the muon sites found in this study, we have calculated the defect formation energy \cite{PhysRevB.78.235104,PhysRevB.81.115311} in order to assess how favorable it is for each of these to be realized (see Table~\ref{table1}). We note that, while the tetrahedral site in rhenium is 0.50~eV higher in energy than the other candidate site discussed below, it is possible that it is occupied nevertheless.  Re atoms in the coordination tetrahedron of the muon can be divided into two distinct environments [Fig.~\ref{fig:fig1}(a)], Re1 at the apex of the tetrahedron and Re2 forming the base of the tetrahedron. Although the presence of the muon distinguishes these different environments, this tetrahedron is not regular, even in the absence of the muon. The muon is not quite at the centre of mass of the coordination tetrahedron [which has fractional coordinates ($\frac{2}{3}$, $\frac{1}{3}$, $\frac{3}{8}$)] but is instead slightly closer to Re1, which results in a larger muon-induced displacement of this Re atom than for the others in the tetrahedron [Fig.~\ref{fig:fig1}(b)]. As is also shown in Fig.~\ref{fig:fig1}(b), the displacements of the nearest-neighbour atoms due to the muon  are small: 0.15~\AA~for Re1 and 0.09~\AA~for Re2. 

\begin{table} 
	\begin{tabular}{lcccccr}
		\hline
		\hline
		Material & $H_{\mathrm{f}}$ & Fractional &   $\sigma_\mathrm{VV}$ & $\sigma_\mathrm{n}$\\
		 & (eV) & coordinates &  (MHz) &  (MHz)\vspace{0.1cm}\\ \hline
		& & & & \vspace{-0.3cm} &\\  
		Re  & $-2.16$ & 0,0,0 &  0.372& 0.314 \cite{PhysRevLett.121.257002}\\
		  & $-1.67$ & $\frac{2}{3}$, $\frac{1}{3}$, 0.388 
		 &  0.420 & \\
		\SRO& $-2.01$ & 0.225, 0.0, 0.184 & 0.048 & 0.02, 0.06 \cite{luke1998}\\
		LaNiC$_{2}$ & $-2.59$ & 0.004, 0.498, 0.112 & 0.120 & 0.08 \cite{PhysRevLett.102.117007}\\
		\spa{} & $-2.21$ & 0.333,   0.662,   0.034    & 0.073 & 0.12 \cite{biswas2013}\\
		& $-1.92$ & 0.009,   0.207,   0.245  & 0.086\\
		Zr$_{3}$Ir & $-3.06$ & 0.001,   0.000,   0.500  & 0.039 & 0.15 \cite{sajilesh2019}\\
		& $-2.72$ & 0.000, 0.000, 0.998 & 0.036&\\
		& $-2.44$ & 0.285, 0.106, 0.223 & 0.033&\\
		& $-2.23$ & 0.573, 0.007, 0.173 & 0.034&\\
		Re$_6$Zr & $-4.98$ & 0.122,	0.120, 0.004 & 0.336 & 0.256 \cite{PhysRevLett.112.107002}\\
        & $-4.92$ & 0.449, 0.001, 0.006 & 0.338&\\	
        & $-4.61$ & 0.504, 0.253, 0.001	& 0.379&\\
		\hline
		\hline
	\end{tabular}
	\caption{Crystallographically distinct muon stopping sites obtained from structural relaxations and their defect formation energies $H_{\mathrm{f}}$. Fractional coordinates are given for the conventional cell. We also show the Van Vleck second moments $\sigma_\mathrm{VV}$ computed for each of the sites (calculated in the limit of strong quadrupolar splitting) and compare these with the measured relaxation rates due to nuclear moments $\sigma_\mathrm{n}$. \label{table1}}
\end{table}

To investigate the possible effects of the muon on the electronic structure of the system, we computed the density of states (DOS) for Re1 and Re2 for the pristine structure and for the structure with a muon at the tetrahedral site, and we show these in Figs.~\ref{fig:fig1}(c) and \ref{fig:fig1}(d), respectively.  The projected density of states (PDOS) of Re1 and Re2 show some small differences due to the symmetry-breaking effect of the implanted muon. However, the changes in the PDOS compared to the pristine system are very minor, particularly in the vicinity of the Fermi energy, where the DOS would have a significant impact on the electronic properties of the system. The PDOS of the muon has the form of a localized state lying around 10~eV below the Fermi energy  and this defect state is therefore unlikely to affect the electronic properties of the system.

In the lowest-energy site in Re, shown in Fig.~S1 in the Supplemental Material (SM) \cite{note1}, the muon is octahedrally coordinated by Re atoms, with Re--$\mu^+$ distances of 2.0~\AA. The displacements due to this site are small, with Re atoms in the coordination octahedron each being displaced by around 0.04~\AA~away from the muon.  These displacements are significantly smaller than those associated with the tetrahedral site. This is likely due to the smaller space available for the muon in a tetrahedral vacancy as compared to an octahedral case, which might also explain the higher total energy of this site.
We also do not observe any significant changes to the electronic structure due to the implanted muon in this case. To compare the computed  sites to the measured spectra \cite{PhysRevLett.121.257002} we computed the expected relaxation rate for each distinct muon stopping site using the Van Vleck second moment in the limit of strong quadrupolar splitting \cite{PhysRevB.20.850}. The calculated relaxation rates are reported in the final column of Table~\ref{table1} and take into account the repulsion of the nearby Re atoms by the muon. We see that both sites give rise to a relaxation rate that is slightly higher than, though broadly consistent with, the value $\sigma=0.314$~MHz observed experimentally. 

\begin{figure}[h]
	\includegraphics[width=\columnwidth]{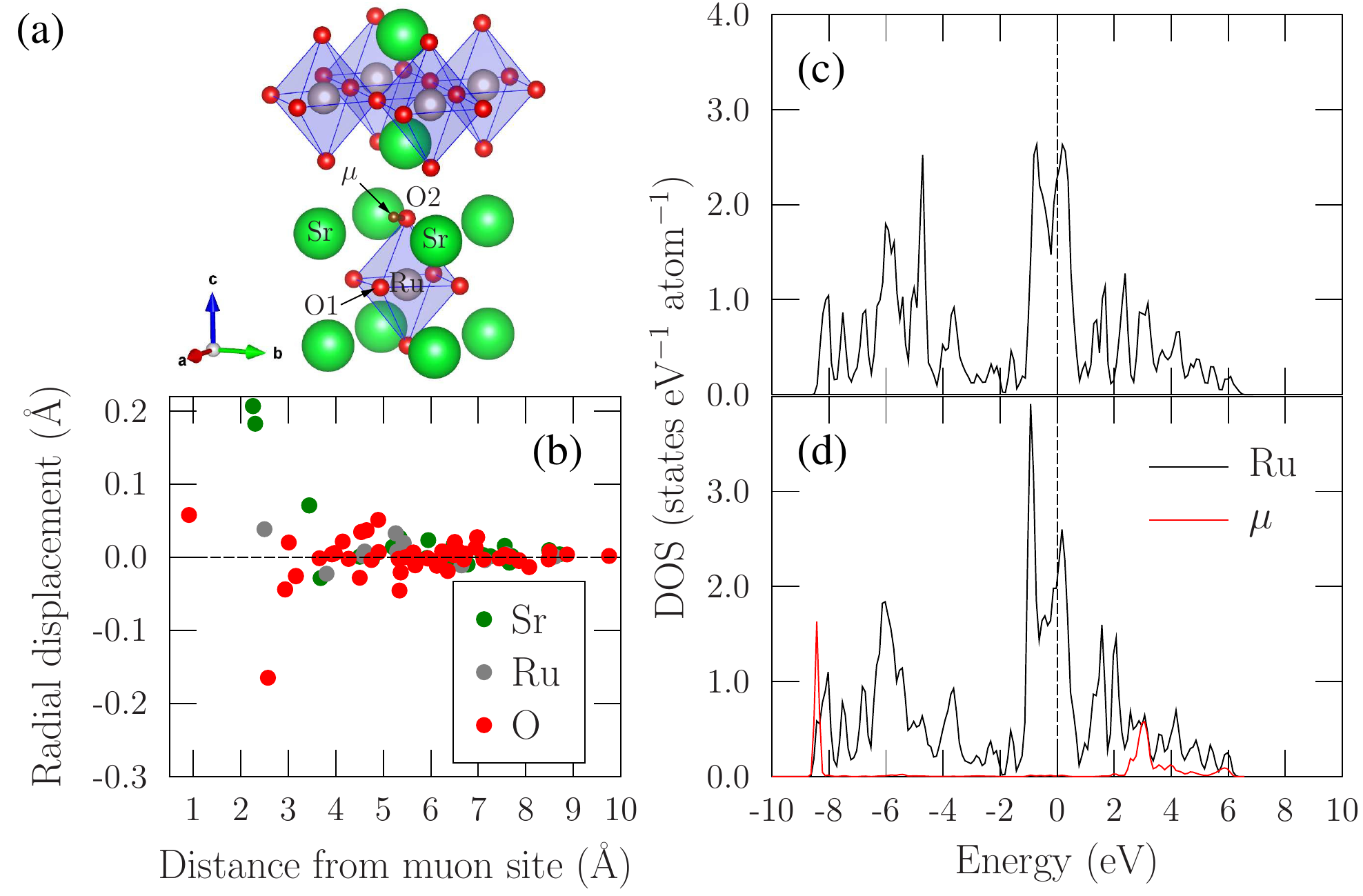}
	\caption{The lowest energy muon site in \SRO{}. (a) The local geometry of the muon site. (b) Radial displacements of atoms as a function of their distances from the muon site. PDOS for the Ru atom closest to the muon site for the structures (c) without a muon and (d) with a muon. Energies are given with respect to the Fermi energy.}
	\label{fig:fig2}
\end{figure}

We now turn to the the layered perovskite superconductor \SRO{}.
In the lowest energy muon site, the muon is  bonded to an oxygen (O2) with bond distance 0.973~\AA, as shown in Fig.~\ref{fig:fig2}(a). This is consistent with  muon sites in other oxides including high-temperature superconducting cuprates~\cite{suter2003,storchak2014} and pyrochlores \cite{foronda2015}, where the muon stops $\approx $1~\AA~from an O anion. The radial displacements of the ions due to the implanted muon are shown in Fig.~\ref{fig:fig2}(b) as a function of their distances from the muon site. The most significant displacements are experienced by the Sr and O1 atoms nearest the muon [labelled in Fig.~~\ref{fig:fig2}(a)], with magnitudes in the range of 0.16 to 0.21~\AA{}. The O2 that forms a bond with the muon is repelled by 0.06~\AA{} away from the muon, while the nearest Ru atom [Ru--$\mu$ distance of 2.54~\AA{}, also indicated in Fig.~\ref{fig:fig2}(a)] experiences a radial displacement of 0.04~\AA{}. For all species, the muon-induced displacement vanishes rapidly as a function of distance from the muon site, such that significant distortions are observed only for atoms within 6~\AA{} of the muon site.

The dominant contribution to the DOS
close to the Fermi energy is that from the Ru atoms \cite{note1}. The effect of muon implantation on the PDOS of Ru atom closest to the muon site is shown  in Figs.~\ref{fig:fig2}(c-d). There is a significant increase in the DOS at around 1~eV below the Fermi energy caused by small changes in the splitting of the Ru 4$d_{zy}$ and 4$d_{zx}$ states at the Fermi level, which are not observed for Ru atoms further away from muon. After summing the $d$-state contributions from all of the Ru ions in the supercell, the small state splitting is no longer resolvable. 
Similar to Re, the PDOS corresponding to the muon lies well below the Fermi energy (around 8~eV below in this case). These results suggest that the implanted muon does not have a significant effect on the local electronic structure of \SRO{}.

\begin{figure}[h]
	\includegraphics[width=\columnwidth]{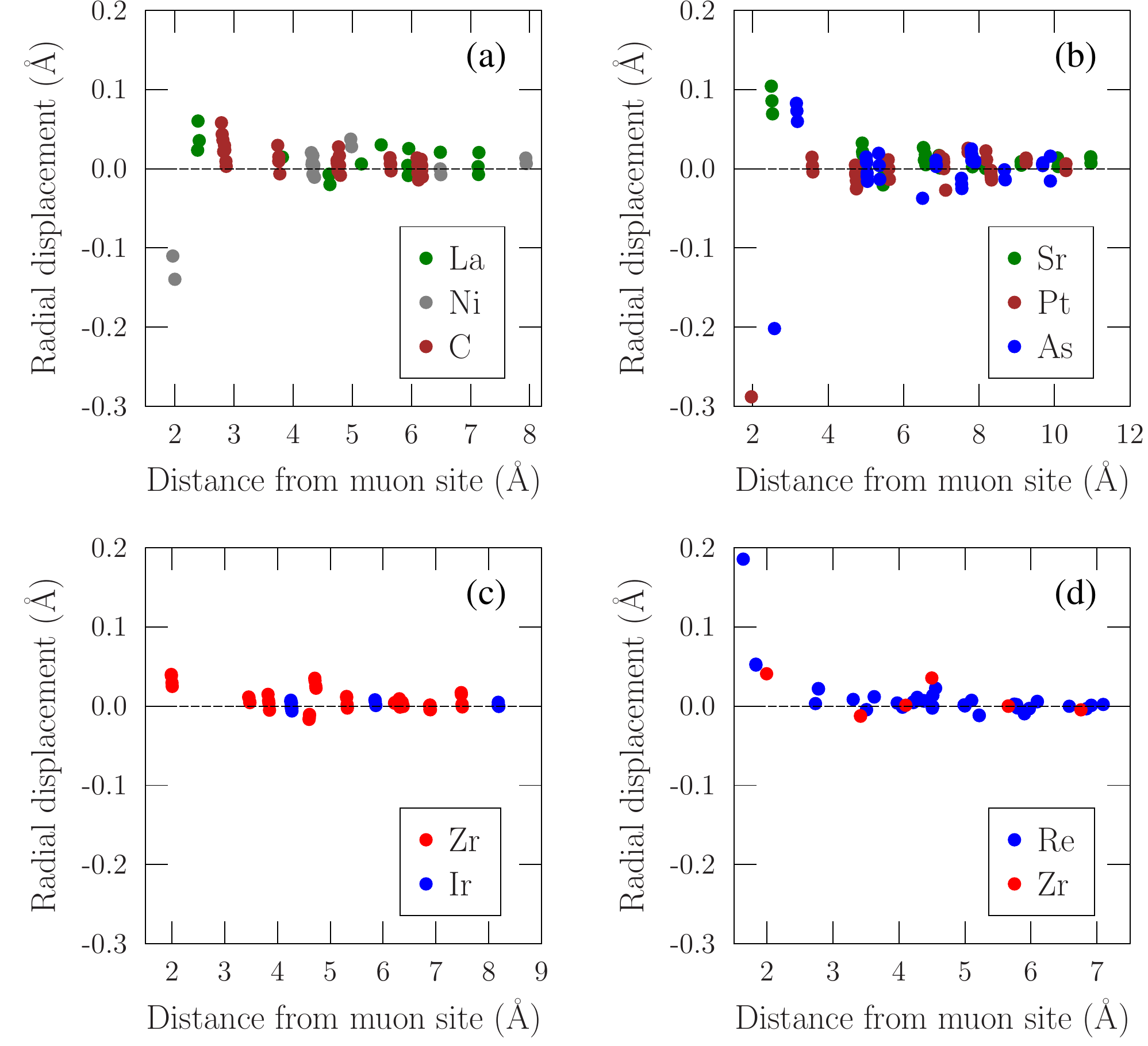}
	\caption{Radial displacements of atoms as a function of their distances from the muon site for the lowest energy muon sites in (a) LaNiC$_2$, (b) SrPtAs, (c) Zr$_3$Ir, and (d) Re$_6$Zr.}
	\label{fig:fig3}
\end{figure}

Muon sites obtained for  other materials in this study are summarized in Table~\ref{table1} and we show the radial displacements of atoms as a function of their distances from the muon for the lowest-energy site in Fig.~\ref{fig:fig3}. We summarize the important features of the muon sites in each of these systems below. (See the Supplemental Material~\cite{note1} for further details.)

For the lowest energy muon site in LaNiC$_2$, the muon is triangularly-coordinated by three La atoms in the $bc$ plane and sits between two Ni atoms along the $a$ axis, with two approximately equal Ni--$\mu^+$ distances of $\approx 1.86$~\AA{}. The Ni atoms are displaced by around 0.11~\AA{} and 0.14~\AA{} towards the muon, as shown in Fig.~\ref{fig:fig3}(a). The La atoms are displaced radially outwards, though by a smaller distances (between 0.02~\AA~and 0.06~\AA). The nuclear relaxation rate computed for the site is very similar to the value $\sigma=0.08$ MHz obtained experimentally \cite{PhysRevLett.102.117007}. 

For SrPtAs, structural relaxations result in three distinct symmetry-inequivalent muon stopping sites. The energy difference of 0.29 eV between the two lowest energy sites (shown in Table~\ref{table1}) is not large enough to rule out all but the lowest energy site and we therefore consider both of these sites as candidate muon sites, and refer to these as site 1 and site 2. Both sites make shorter bond distances with Pt than with Sr and As atoms. As seen in Fig.~\ref{fig:fig3}(b), for site 1, the  Pt atom nearest the muon experiences the largest displacement, with magnitude 0.3 \AA{}. For site 2, the displacement of the nearest Pt atom is much smaller ($\approx 0.12$~\AA), but significant displacements persist to larger distances away from the muon site \cite{note1}.

For \zri{}, structural relaxations result in 10 distinct muon sites, which we cluster into four groups by considering the proximity of their positions within the unit cell, and we include a representative member of each group in Table~\ref{table1}.  In the lowest energy site, the muon is tetrahedrally coordinated by Zr atom, with Zr--$\mu$ bond lengths of 2.03 \AA{}. For this site, the muon does not introduce any significant distortions to its host, with all displacements being 0.04 \AA{} or smaller, as seen in Fig.~\ref{fig:fig3}(c). In fact, for all sites in \zri{}, the maximum displacement remains below 0.1 \AA{}.

For the Re-Zr alloy Re$_6$Zr, obtaining the precise Re:Zr composition would require a prohibitively large supercell, so we instead considered compositions Re$_{49}$Zr$_{9}$ and Re$_{50}$Zr$_{8}$ as close approximations. The distinct sites for Re$_{50}$Zr$_{8}$ are reported in Table~\ref{table1}. For the lowest energy site, the muon is tetrahedrally coordinated by 3 Re atoms and 1 Zr atom, with Re--$\mu$ distances of 1.83 \AA{}, 1.88 \AA{} and 1.88 \AA{} and a Zr--$\mu$ distance of 2.04 \AA{}. The coordination tetrahedron of the muon in this site (and for other tetrahedral sites in this system) is therefore highly irregular, with this being the case even before the addition of a muon. As seen in Fig.~\ref{fig:fig3}(d), the displacement of each of the ions in the coordination tetrahedron reflects their proximity to the muon site; atoms that are closer to the muon are displaced by a greater amount, with the displacement of the Re atom closest to the muon being significantly larger than for any of the other atoms in the system.

There are no significant changes to the DOS in the vicinity of the Fermi energy as a result of implanting a muon in any of the materials studied. The significant contributions from muon PDOS typically lie between 6 and 10~eV below the Fermi energy, and the PDOS corresponding to the other species remains basically unchanged near the Fermi energy. While we often find evidence for some hybridization between the muon states and the states of ions in the system, the fact that these muon states lie so far below the Fermi energy means that this is unlikely to affect the electronic properties of the system, even locally. We have found that the electronic behavior of the muon in these systems is very similar to that in the conventional superconductor Nb \cite{note1}, with no notable differences that could be responsible for muon-induced internal fields.

Another possible way in which the muon could perturb its host is by altering the charge states of nearby atoms. For \zri{}, it was found that, in the vicinity of the muon site, the occupation of the $d$ states and charges of the neighbouring Zr atoms change. This was particularly true for the lowest energy site, which has four nearest-neighbour Zr atoms. This is due to the redistribution of the charge on the atoms bonding with the muon, since the charge on the muon remains the same for calculations with muonium as it does for $\mu^+$. In fact, the muon exists in a charge state resembling muonium in all of materials in this study, which is consistent with the fact that the muon PDOS lies well below the Fermi energy and hence these states would be expected to be occupied. In general, the overall (+1) charge of the unit cell is maintained by the having the charges on all of the other atoms in the host system become very slightly more positive, rather than through significant changes in the charge state of particular atoms.

We also carried out a series of spin-polarized calculations to investigate the possibility of muon-induced spin density which could, in principle, act as a source of a non-zero local magnetic field. For LaNiC$_2$, we found no appreciable spin density, both for the pristine structure and for the structure including an implanted muon ($<0.01 \hbar/2$ per atom according to a Mulliken analysis). Similar spin-polarized calculations on \zri{} do not show the presence of spin density at the muon site. After muon implantation, the the small  (0.0122~a.u.) spin density on each of the Zr atoms is altered (by 0.004~a.u.) and vanishes for the nearest-neighbor Zr atoms, confirming that the muon does not induce magnetization. It is sometimes possible to form local moments on impurities in metals through their resonant interaction with conduction electrons \cite{ziman_1972,coleman_2015}. However, the facts that the muon is a light impurity and that the muon PDOS lies so far below the Fermi energy makes moment formation on the muon extremely unlikely here (a criterion for this is provided in the SM \cite{note1}).

In summary, we have carried out a systematic investigation of the muon sites in a variety of superconducting compounds which purportedly exhibit TRSB in order to assess the extent of any muon-induced effect that could, in itself, give rise to the observation of a spontaneous field.   
Because the muon acts like a charged impurity, the most significant effects it could have are (i) on the local structural arrangement of atoms or ions close to the muon  and (ii) on the local electronic
structure. 
For point (i), our results show that in all cases studied
the structural distortion involves only a modest alteration in the  positions of nuclei
that is rapidly suppressed with distance. Point (ii) is potentially more important since many superconductors that are candidates for TRSB have 
several bands crossing
the Fermi energy, leading to multiple Fermi-surface
sheets; if the muon were to appreciably alter the electronic
structure near $E_{\mathrm{F}}$ this might conceivably provide a mechanism for the muon to couple to some muon-induced spin density (although one would then need an explanation for why this effect tracks the order parameter in the superconducting state).  However, our results show that in these materials the changes to the local electronic structure resulting from muon implantation 
occur several eV below $E_{\rm F}$, well away from the superconducting gap (which is a few meV around $E_{\rm F}$), precluding any direct effect of the muon on the local superconducting state.  This contrasts with results on hydrogenic impurity states in several semiconductors (e.g.\ ZnO \cite{ZnO1,ZnO2} and HfO \cite{HfO}) 
in which the muon level is found within the gap and close to $E_{\rm F}$ \cite{cox_2009}.
Moreover, since the muon level is deep below $E_{\rm F}$ in all the compounds considered in this paper, it acts as a neutral defect and so the only perturbation of the local charge density is caused by the (very small) movement of the nearby ions that drag their charge density with them.  In these systems, we find that the calculated change in the electronic charge on nearby ions is typically $<0.4$\%.  This puts a tight constraint on any models which attempt to explain the spontaneous fields as being due to the suppression of the superconducting order parameter by an imagined screening cloud of charge density around the muon \cite{miyake2017,miyake2019}.  The calculations show that in these systems the muon is instead a rather benign defect that produces minimal effect on the local charge density.
We therefore conclude that the observation of spontaneous local fields in superconductors exhibiting TRSB is an effect which is intrinsic to these compounds and not a result of a muon-induced effect.

Finally, we note that the techniques demonstrated here are applicable well beyond the
question of muons in superconductors exhibiting TRSB.
These results suggest such that systematic calculations of muon sites in materials is a promising, and necessary, means to assess the influence of the stopped muon on any exotic physics for which it is being used to act as an experimental probe.

 Research data from the UK effort will be made available via Ref.~\cite{data}.

This work is supported by EPSRC (UK), under
Grants No. EP/N024028/1 and No. EP/N032128/1. We
acknowledge computing resources provided by STFC Scientific Computing Department’s SCARF cluster, Durham
Hamilton HPC, and the ARCHER UK National Supercomputing Service. We also acknowledge computing resources provided by CINECA under Project ID IsC58 and the HPC resources at the University of Parma, Italy. I.J.O.\ acknowledges funding from the SUPER (Supercomputing Unified Platform - Emilia-Romagna) regional project. B.M.H.\ and I.J.O.\ contributed equally to this work.

\end{document}


\title{Supplemental Material for ``Intrinsic nature of spontaneous magnetic fields in superconductors with time-reversal symmetry breaking"} 

\author{B. M. Huddart}
\affiliation{Centre for Materials Physics, Durham University, Durham DH1 3LE, United Kingdom}
\author{I. J. Onuorah}
\affiliation{Department of Mathematical, Physical and Computer Sciences, University of Parma, 43124 Parma, Italy}
\author{M. M. Isah}
\affiliation{Department of Mathematical, Physical and Computer Sciences, University of Parma, 43124 Parma, Italy}
\author{P. Bonf\`{a}}
\affiliation{Department of Mathematical, Physical and Computer Sciences, University of Parma, 43124 Parma, Italy}
\author{S. J. Blundell}
\affiliation{Department of Physics, Clarendon Laboratory, Oxford University, Parks Road, Oxford OX1 3PU, United Kingdom}
\author{S. J. Clark}
\affiliation{Centre for Materials Physics, Durham University, Durham DH1 3LE, United Kingdom}
\author{R. De Renzi}
\affiliation{Department of Mathematical, Physical and Computer Sciences, University of Parma, 43124 Parma, Italy}
\author{T. Lancaster}
\affiliation{Centre for Materials Physics, Durham University, Durham DH1 3LE, United Kingdom}

\date{\today}

\maketitle

\section{\label{sec:methods}Methods}
\subsection{Density functional theory}
To identify the muon stopping sites and to evaluate the effect of the implanted muon on the host system, we carried out a suite of plane-wave basis-set density functional theory (DFT) calculations. Calculations on rhenium, LaNiC$_2$ and Re$_6$Zr were carried out using \textsc{castep} \cite{CASTEP}, whereas calculations on Sr$_2$RuO$_4$, SrPtAs and Zr$_3$Ir were carried out using {\sc Quantum ESPRESSO} \cite{QE-2009,QE-2017}. In all cases, we work within the generalized-gradient approximation (GGA) using the PBE functional \cite{pbe1996}. The system is treated as non-spin-polarized, unless specified otherwise. In this section we outline the series of calculations that were applied to each of the materials in this study.

The first step of our procedure is to obtain the muon stopping sites using a structural relaxation approach. In this approach, initial structures comprising the material and an implanted muon (represented by a hydrogen pseudopotential) are allowed to relax until the changes in the total energy and forces are fall below a convergence threshold. For all of our calculation carried out using \textsc{castep}, structures are allowed to relax until forces on the atoms were all $<5\times 10^{-2}$ eV \AA$^{-1}$ and the total energy
and atomic positions converged to $2\times10^{-5}$ eV per atom and $1\times10^{-3}$~\AA, respectively. For calculations using \textsc{quantum espresso} the atomic relaxations through force and energy minimization were performed with the threshold for the forces  and total energy set to 10$^{-3}$ Ry/a.u and 10$^{-4}$ Ry, respectively. 
This approach requires the use of supercells in order to minimise the spurious self-interaction of the muon that results from the periodic boundary conditions. Initial muon positions can be either inferred from the minima of the electrostatic potential for the host crystal (which is likely to be a stable position for the positive $\mu^+$) or obtained from by sampling the space within the unit cell. The symmetry of the host crystal can be used to reduce the number of initial positions that need to be sampled. Low-energy relaxed structures provide candidate muon stopping sites and also allow us to assess the significance of any structural distortions induced by the muon on its local environment.

After obtaining the muon site and assessing whether the presence of the muon leads to significant structural distortions we proceed to study its effect on the local \textit{electronic} structure. We do this by calculating the projected density of states (PDOS) for the system for supercells with or without the muon. Projections were made onto each of the atoms, each of the angular momentum channels, and, in select cases, according to the magnetic quantum number.  This allows us to see where the electronic state corresponding to the muon sits relative to the Fermi energy and also to assess the possible impact of the muon on the DOS of the other species near the Fermi energy.

\subsection{Defect formation energies}
Defect formation energies, $H_\mathrm{f}$, are commonly calculated for semiconductors using DFT \cite{PhysRevB.78.235104,PhysRevB.81.115311}, as obtaining this quantity enables the calculation of important properties such as defect concentrations. Here, we use this formalism to compare the energetics of muon sites across the various materials included in our study. The defect formation energy of a defect of charge $q$ is given by
\begin{equation}
    H_\mathrm{f}=E_{\mathrm{D},q}-E_\mathrm{H}+qE_\mathrm{F}+\sum_{\alpha}n_{\alpha}\mu_{\alpha},
\end{equation}
where $E_{\mathrm{D},q}$ and $E_\mathrm{H}$ are the total energies of the host+defect supercells and host-only supercells respectively, $E_\mathrm{F}$ is the Fermi energy of the host cell, and $n_\alpha$ is the number of atoms of species $\alpha$ {\it removed}, with reference chemical potential $\mu_\alpha$. We have calculated $H_\mathrm{f}$ for the addition of $\mu^+$, for which $q=+1$ and $n_\alpha=-1$. We take the reference state of $\mu^+$ to be muonium, which is chemically equivalent to a hydrogen atom. We therefore use $\mu_\alpha=-13.604$~eV, obtained from a \textsc{castep} calculation on an isolated hydrogen atom. The resulting values of $H_\mathrm{f}$ for select sites are given in Table~1 in the main text. Note that, within each material, the differences in defect formation energies are solely due to differences in $E_{\mathrm{D},q}$. However, the defect formation energy allows us to compare the energetics of sites belonging to different materials, which cannot be done using total energies of the host+defect supercells alone.

\subsection{Van Vleck relaxation rates}
To reconcile the calculated muon sites with the measured spectra we computed the expected relaxation rate for the site from the the van Vleck second moment in the limit of strong quadrupolar splitting \cite{PhysRevB.20.850}. In a zero-field experiment this formalism predicts a relaxation rate given by
\begin{equation}\label{vv}
    \sigma^2_\mathrm{VV}=\frac{4}{9}\sum_iB_iQ(I),
\end{equation}
with
\begin{equation}
B_i=\left(\frac{\mu_0}{4\pi}\right)^2I(I+1)\gamma^2_\mu\gamma^2_I\hbar^2r^{-6}_i,    
\end{equation}
where the nucleus at site $i$ at a distance $r_i$ from the muon site has nuclear spin $I$ and a nuclear gyromagnetic ratio $\gamma_I$. The nuclear gyromagnetic ratio $\gamma_I=g\mu_\mathrm{N}/\hbar$, where $\mu_\mathrm{N}$ is the nuclear magneton and the $g$-factor is related to the nuclear magnetic moment $\mu$ via $g=\mu/I$. The factor $Q(I)=1$ for integer spin $I$, but for half-integer spin is instead given by
\begin{equation}
    Q(I)=1+\frac{3(I+1/2)}{8I(I+1)}.
\end{equation}

Because the expression for $\sigma^2$ in Eq.~\eqref{vv} is a linear combination of the contributions from each of the nuclei in the system, we can include the effects of the muon-induced distortions on the relaxation rate due to changes in $r_i$ by calculating
\begin{equation}
    \sigma^2_\mathrm{full,dist}=\left[\sigma^2_\mathrm{cell,dist}-\sigma^2_\mathrm{cell,undist}\right]+\sigma^2_\mathrm{full,undist}.
\end{equation}
The relaxation rate $\sigma^2_\mathrm{cell,dist}$ is obtained using the ionic positions from the relaxed cell including the muon, whereas $\sigma^2_\mathrm{cell,undist}$ is calculated from a cell of the same size that does not include a the distortions associated with the implanted muon. It is assumed that the significant displacements do not extend beyond the boundaries of the simulation cell and hence the contribution from nuclei at larger distances from the muon site are calculated from an undistorted structure comprising several unit cells in each direction to obtain $\sigma^2_\mathrm{full,undist}$. Note that this relaxation rate converges very quickly with system size due to the factor of $r^{-6}_i$ in $B_i$.  

The linearity of Eq.~\eqref{vv} also allows us to straightforwardly deal with different isotopes. Consider a nucleus of species $i$ with multiple isotopes $\alpha$ each having nuclear spins $I_\alpha$, moments $\mu_\alpha$, and natural abundances $f_\alpha$. Experimentally, each muon at a given crystallographically distinct site would see one of many possible configurations of isotopes. However, for an ensemble containing a large number of muons, the linearity of Eq.~\eqref{vv} makes the summation over all possible configurations very straightforward, yielding the result
\begin{equation}
    \sigma^2_\mathrm{VV}=\frac{4}{9}\sum_{i,\alpha} f_\alpha B_{i,\alpha}Q(I_\alpha).
    \end{equation}
    
\section{Calculations and muon site details}

Below we give parameters and computational methods used in each of the computations discussed in the main paper. We also provide some further details of the muon position and computed electronic structure specific to each of the materials. 

\subsection{\label{sec:re}Rhenium}

Rhenium crystallizes in the centrosymmetric hexagonal
 $P$6$_3$/$mmc$ space group with $a = b = 2.762$~\AA~and $c = 4.457$~\AA~\cite{PhysRevLett.121.257002}. We used a plane-wave cutoff energy of 600 eV and a $15 \times 15 \times 8$ Monkhorst-Pack grid \cite{mpgrid1976} for Brillouin zone integration, resulting in total energies that converge to 0.002 eV per cell.  The unit cell was allowed to relax and we obtain optimised lattice parameters $a = b = 2.777$~\AA~and $c = 4.478$~\AA, which are within 0.5\% of the experimental values. We used the DFT-optimized lattice parameters and ionic positions in all subsequent calculations.

Structural relaxations were carried out on a supercell comprising $3 \times 3\times2$ conventional unit cells of Re to reduce the unphysical interaction of the muon and its periodic images.  Due to the enlarged unit cell, we instead used a  $5 \times 5 \times 4$ Monkhorst-Pack grid \cite{mpgrid1976} for these calculations.
Initial structures comprising a muon and the Re supercell were generated by requiring the muon to be at least 0.25~\AA~away from each of the muons in the other structures generated (including their symmetry equivalent positions) and at least 1.0~\AA~away from any of the atoms in the cell.  This resulted in 19 structures which were subsequently allowed to relax.  

\begin{figure}[h]
	\includegraphics[width=\columnwidth]{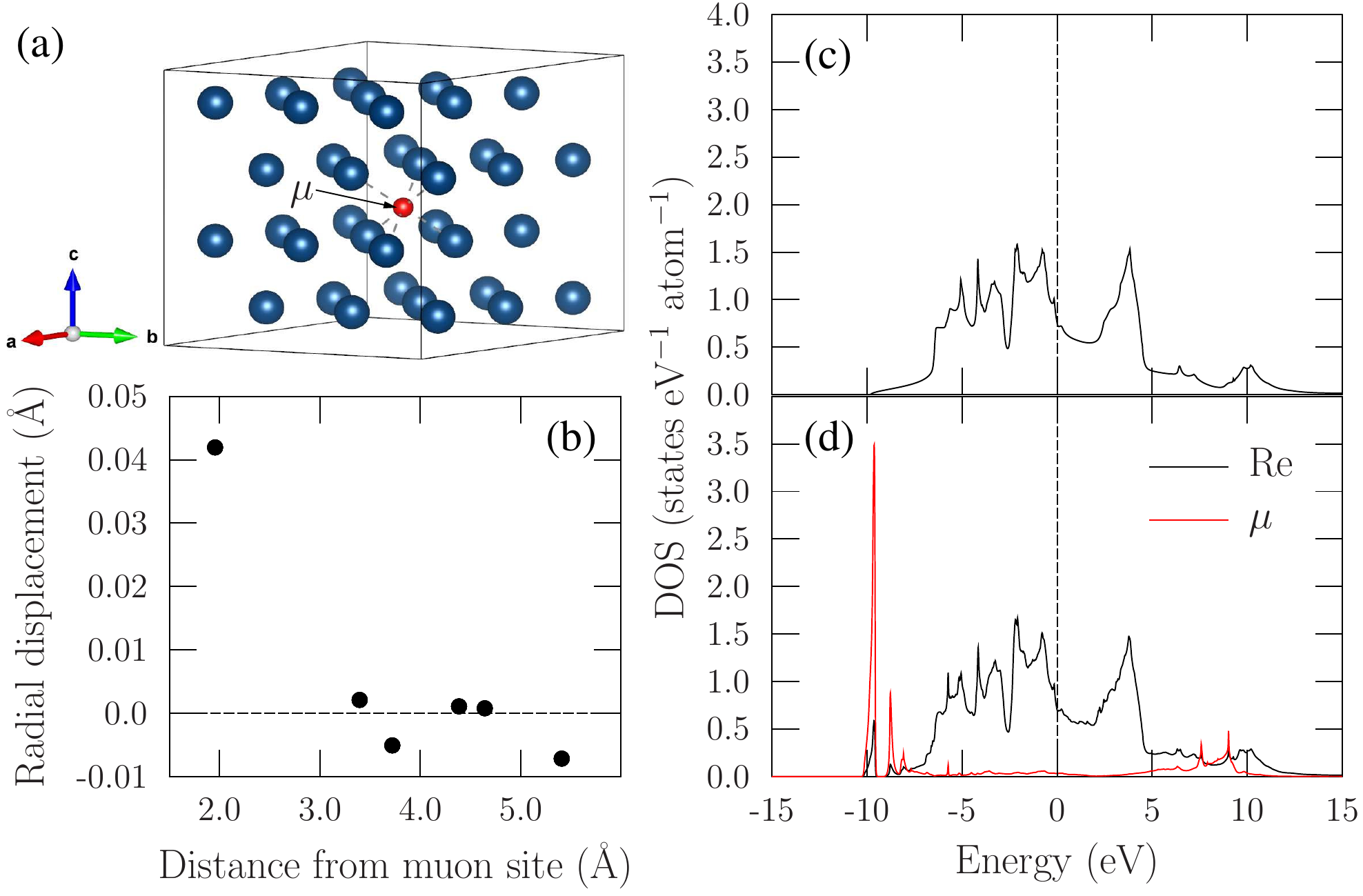}
	\caption{The lowest energy muon site in elemental rhenium. (a) The muon is octahedrally coordinated by Re atoms. (b) Radial displacements of the Re atoms as a function of their distances from the muon site. Projected density of states (PDOS) for the nearest-neighbor Re atoms (c) without a muon and (d) with a muon.}
	\label{fig:fig1}
\end{figure}

These structural relaxations yield two crystallographically distinct muon stopping sites, summarized in Table.~1 in the main text. 
In the lowest energy site [Fig.~\ref{fig:fig1}(a)] the muon is octahedrally coordinated by Re atoms, with Re--$\mu^+$ distances of 2.0 \AA. The displacements due to this site are small, with Re atoms in the coordination octahedron each being repelled by around 0.04~\AA~away from the muon. We find a second crystallographically distinct site, 0.50 eV higher in energy, where the muon is instead tetrahedrally coordinated by Re atoms [see Fig.~1(a) in the main text]. The Re--Re distances along the edges of the base are 2.78 \AA, whereas the distance between a vertex in the base and the apex is 2.75 \AA. In the presence of a muon, the Re atoms in the base of the tetrahedron are repelled by 0.09 \AA, whereas the Re atom at the apex is repelled by 0.15 \AA. The muon is closer to the apical Re (1.77 \AA) than it is to the Re atoms in the base of the tetrahedron (1.81 \AA).


To investigate the possible effects of the implanted muon on the electronic structure of the system, we computed the density of states (DOS) with and without the muon, for both crystallographically distinct muon sites. We used a finer $15 \times 15 \times 12$ Monkhorst-Pack grid \cite{mpgrid1976} for $k$-point sampling in these calculations. For the octahedral site, we show the projected density of states (PDOS) for each of the species in the system for without and with an implanted muon in Figs.~\ref{fig:fig1}(c) and ~\ref{fig:fig1}(d) respectively. We see that the DOS does not change significantly around the Fermi energy with the density of states due to the muon lying around $10$ eV below the Fermi energy. This is also true for the case of the tetrahedral site, as shown in Figs.~1(c) and 1(d) in the main text. 

\begin{figure}
 \centering
\includegraphics[width=0.9\columnwidth]{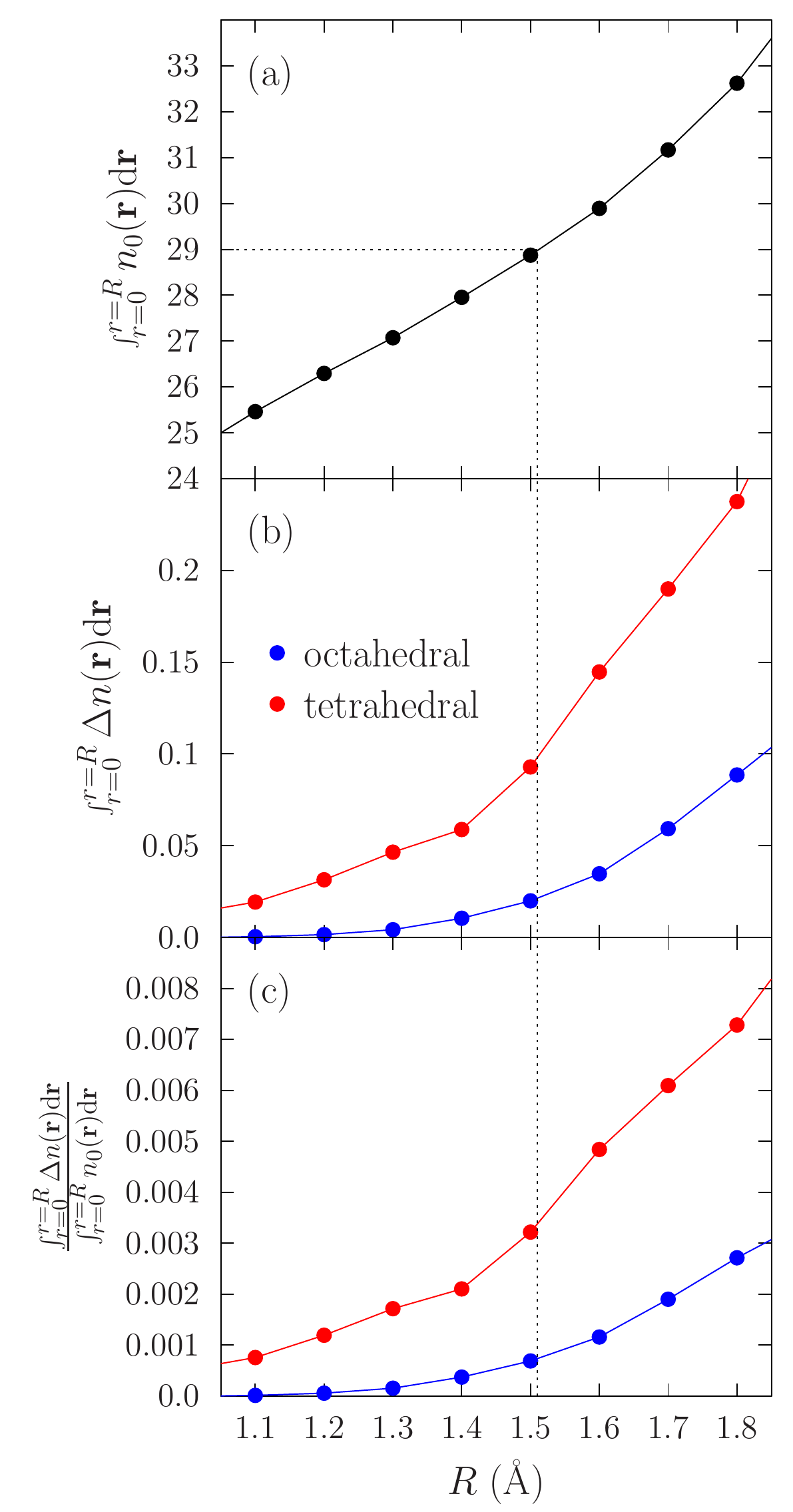}
 \centering
\caption{\label{fig:re_charge_den} Electron densities in rhenium integrated over spheres of radii $R$. (a) Charge density $n_0$ for the pristine system. (b) Charge density difference $\Delta n$ between the system with and without a muon. (c) Charge density difference as a fraction of $n_0$.}
\end{figure}

Finally, we examined the effect of the implanted muon on the charge states of the atoms in the host for both the octahedral and tetrahedral sites. This was done by integrating the charge (electron) density over spheres of various radii $R$ centered on a nearest-neighbour Re atom for the system with and without a muon. (For the tetrahedral site, the Re atom at the apex of the tetrahedron was chosen, as this is the closest to the muon.) The integral of the charge density $n_0(\textbf{r})$ for the pristine structure as a function of $R$ is shown in Fig.~\ref{fig:re_charge_den}(a) and counts the number of valence electrons enclosed by the sphere (i.e. those electrons not removed through the use of pseudopotentials). The integrated difference in charge densities for each site $\Delta n=n_\mu-n_0$, where $n_\mu$ is the electron density for the system containing a muon, is shown in Fig.~\ref{fig:re_charge_den}(b). Each Re atom has 29 valence electrons, which are accounted for when using a radius of integration $r \approx 1.5$~\AA. At this radius, the sphere of integration for the system with a muon contains 0.02 more or 0.07 more electrons than for the case of the pristine system, for the octahedral and tetrahedral sites, respectively. This extra electron density is associated with the muon, rather than the Re atoms. A Mulliken population analysis instead shows that the nearest-neighbour Re atoms become slightly positive in both cases, with induced charges of +0.06$e$ and +0.08$e$ for the octahedral and tetrahedral sites, respectively.  However, both approaches find that the magnitude of the change in the charge states is very small, as seen in Fig.~\ref{fig:re_charge_den}(c), where the charge difference is expressed as a fraction of the total number of valence electrons enclosed by the sphere. The changes in the number of valence electrons are less than 0.4\% of the initial number of electrons.

\subsection{\label{sec:sro}$\textrm{Sr}_{2}\textrm{RuO}_{4}$}
\begin{figure}
 \centering
\includegraphics[width=\columnwidth]{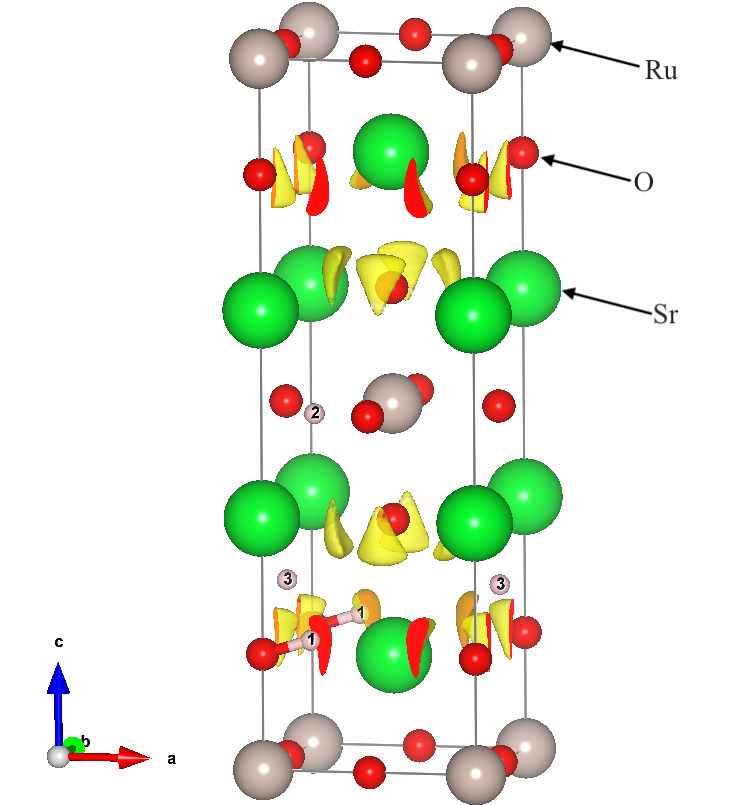}
 \centering
\caption[\SRO{} structure]{\label{fig:sroa} Muon stopping sites in the \SRO{} unit cell, with isosurfaces of the electrostatic potential, visualized using \textsc{vesta}~\cite{vesta2008}. }
\end{figure}

\SRO{} crystallizes in the centrosymmetric $I4/mmm$ space group. In this system, O$_4$ atoms form an octahedral at the centre of a cube defined by Sr$_2$  at the corners, while the Ru cation is at the center of the octahedral (see Fig.~\ref{fig:sroa}). The unit cell has lattice parameters $a=b=$3.871 \AA{} and $c = 12.702$ \AA{}, with Sr atoms occupying the 4e Wyckoff positions with fractional coordiantes (0, 0, 0.3538),  Ru atoms at 2a with (0.0, 0.0, 0.0), one of the oxygen atoms (O1) at 4c with (0.0, 0.5, 0.0) and the other (O2) at 4e with (0.0, 0.0, 0.1630). In our calculations, the potential at the core was approximated using the optimized Norm-Conserving Vanderbilt (ONCV)\cite{hamann2013} pseudopotential for Ru atoms and ultrasoft~\cite{ultsft1990,gbrv2014} pseudopotential for the Sr and O atoms. The pseudopotential choices were made to allow for correct description of the 4$d$ behaviour in Ru and to allow convergence of the self-consistent iteration.  The cut-off for the plane waves and the charge density used are 70 Ry and 700 Ry respectively.

The trial starting positions for the muon are a grid of uniform positions, reduced by the space group symmetry to 24 inequivalent starting positions. Each of the trial starting muon position was modelled in a 3$\times$3$\times$1 supercell, containing 126 atoms. 
 A $3 \times 3 \times 3$ Monkhorst-Pack grid of $k$-points was used for the Brillouin zone sampling. The 24 final muon positions collapse into 3 inequivalent candidate positions when the total energies are considered, with all of the muon positions within each cluster being symmetry-equivalent. We report a representative candidate muon site belonging to each of the clusters in Table~\ref{tab:MuonSiteDFTsro}.


\begin{table}[!htbp]
	\centering
	\caption{Fractional coordinates and energies (relative to the lowest energy site) for the candidate muon stopping sites in \SRO{}.}
	\label{tab:MuonSiteDFTsro}
		\begin{tabular}{lcr}
		\hline
		\hline
	        Label & Site Position & $\Delta E$ (eV)\\ \hline
			A  & (0.225, 0.0, 0.184) & 0 \\
			B  & (0.233, 0.071, 0.5) & 0.97  \\
			C & (0.0, 0.498, 0.249) & 1.61 \\ 
			\hline
			\hline		
		\end{tabular}
\end{table}

Muon site A is significantly lower in energy than sites B and C and is therefore likely the sole stopping site. The energy difference between A and B ($\approx 1$ eV) is a conservative lower limit to the potential well depth for the muon in the A site. Since this is much larger that the typical muon zero point energy (0.5 eV), the quantum nature of the muon does not need to be considered here. Site A is shown in Fig.~2(a) in the main text, with the displacements of the Sr, Ru and O atoms as a function of their distance from the muon site shown in Fig.~2(b). 

\begin{figure}
\centering
\includegraphics[width=\columnwidth]{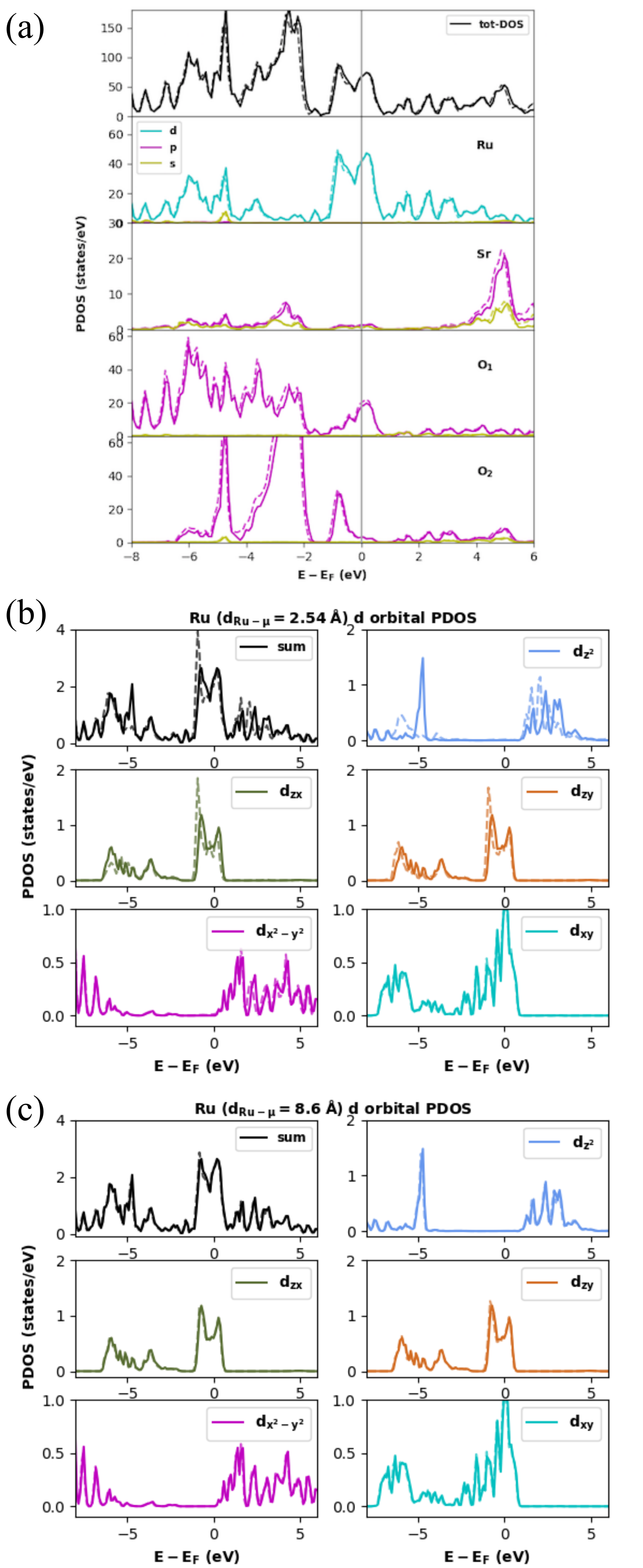}
\caption[PDOS of \SRO{}]{\label{fig:pdossroo} Comparison of the PDOS for the supercells with the muon (solid lines) and those without the muon (dotted lines) (a) The total and partial projected density of states (PDOS) of \SRO{}. The dotted lines are the PDOS of the supercell with the muon.   (b) PDOS of Ru 4$d$  orbital contribution,  for Ru--$\mu$ distance of 2.54~\AA{} (c) PDOS of Ru 4$d$ orbital contribution with the muon, for Ru--$\mu$ distance of 8.6~\AA{}. The Fermi energy has been shifted to 0 eV.}
\end{figure}

We considered further the effect of the implanted muon and its associated lattice distortion on the electronic structure by calculating its effect on the density of states (DOS). The DOS of \SRO{} is that of  a metal (similar to results in Ref.~\cite{hao2014}) and the system remains metallic after the muon implantation [Fig.~\ref{fig:pdossroo}(a)]. To understand the orbital contribution to the DOS, a projection of the DOS to the different atomic orbitals was performed. Very close to the Fermi energy, the dominant contribution to the DOS are the Ru partial density of states (PDOS) with 4$d$ character [see Fig.~\ref{fig:pdossroo}(a)], similar to LDA results in Ref.~\cite{oguchi1995,singh1995}. We also see a hybridization between the Ru 4$d$ states and O 2$p$ states near the Fermi energy. These contributions are similar and do not show any significant changes after muon implantation.

Further analysis of the 4$d$ states show that the conduction holes are mostly contributed by the Ru 4$d_{zy}$, 4 $d_{zx}$ and 4$d_{xy}$ PDOS [see Figs.~\ref{fig:pdossroo}(b) and \ref{fig:pdossroo}(c)]. This is consistent with results obtained using the near-edge x-ray absorption and photoemission spectroscopy \cite{schmidt1996}. The effect of an implanted muon on the PDOS is demonstrated in Figs.~\ref{fig:pdossroo}(b) and \ref{fig:pdossroo}(c), where the dotted lines represent the 4$d$ states contributions due to Ru atoms closest to (2.54 \AA{}) and far from (8.6 \AA{}) the muon site, respectively.  Small changes in the splitting of the states at the Fermi level are observed for the 4$d_{zy}$ and 4$d_{zx}$ states of the Ru atom close to the muon [Fig.~\ref{fig:pdossroo}(b)], but are not observed for the Ru atom far from the muon [Figs.~\ref{fig:pdossroo}(c)]. After summing all of the $d$-state contributions of the Ru ions in the supercell, the small state splitting is no longer significant. 

\begin{figure}
 \centering
\includegraphics[width=\columnwidth]{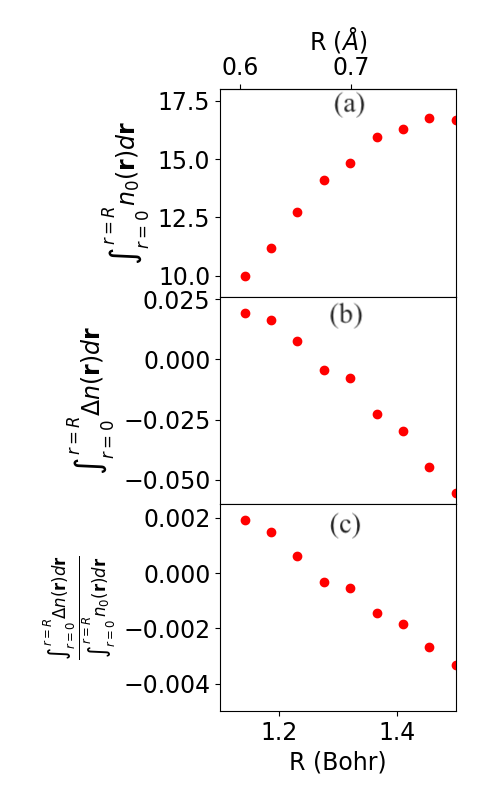}
 \centering
\caption{\label{fig:sro_charge_den} Electron densities at the Ru atom closest to the muon site in Sr$_2$RuO$_4$, integrated over spheres of radii $R$. (a) Charge density $n_0$ for the pristine system. (b) Charge density difference $\Delta n$ between the system with and without a muon. (c) Charge density difference as a fraction of $n_0$.}
\end{figure}

We investigated the effect of the implanted muon on the electron density in Sr$_2$RuO$_4$ using a similar approach as was used for rhenium. The spheres of integration were centered on Ru atom nearest the muon and the integrated charge densities as a function of radius are shown in Fig.~\ref{fig:sro_charge_den}. Ru has 16 valence electrons and, as seen in Fig.~\ref{fig:sro_charge_den}(a), these are accounted for by integrating over a sphere with a radius of around 1.4 Bohr. For a sphere of this radius, 0.03 fewer electrons are enclosed by the sphere for the system including a muon than for the system without, as shown in Fig.~\ref{fig:sro_charge_den}(b). The Ru atom is therefore slightly more positive due to the presence of an implanted muon. However, as seen in Fig.~\ref{fig:sro_charge_den}(c), this corresponds to a decrease of just 0.2\% in the number of valence electrons in the vicinity of Ru.

\subsection{\label{sec:lanic2}LaNiC$_2$}
\begin{figure}[h]
	\includegraphics[width=\columnwidth]{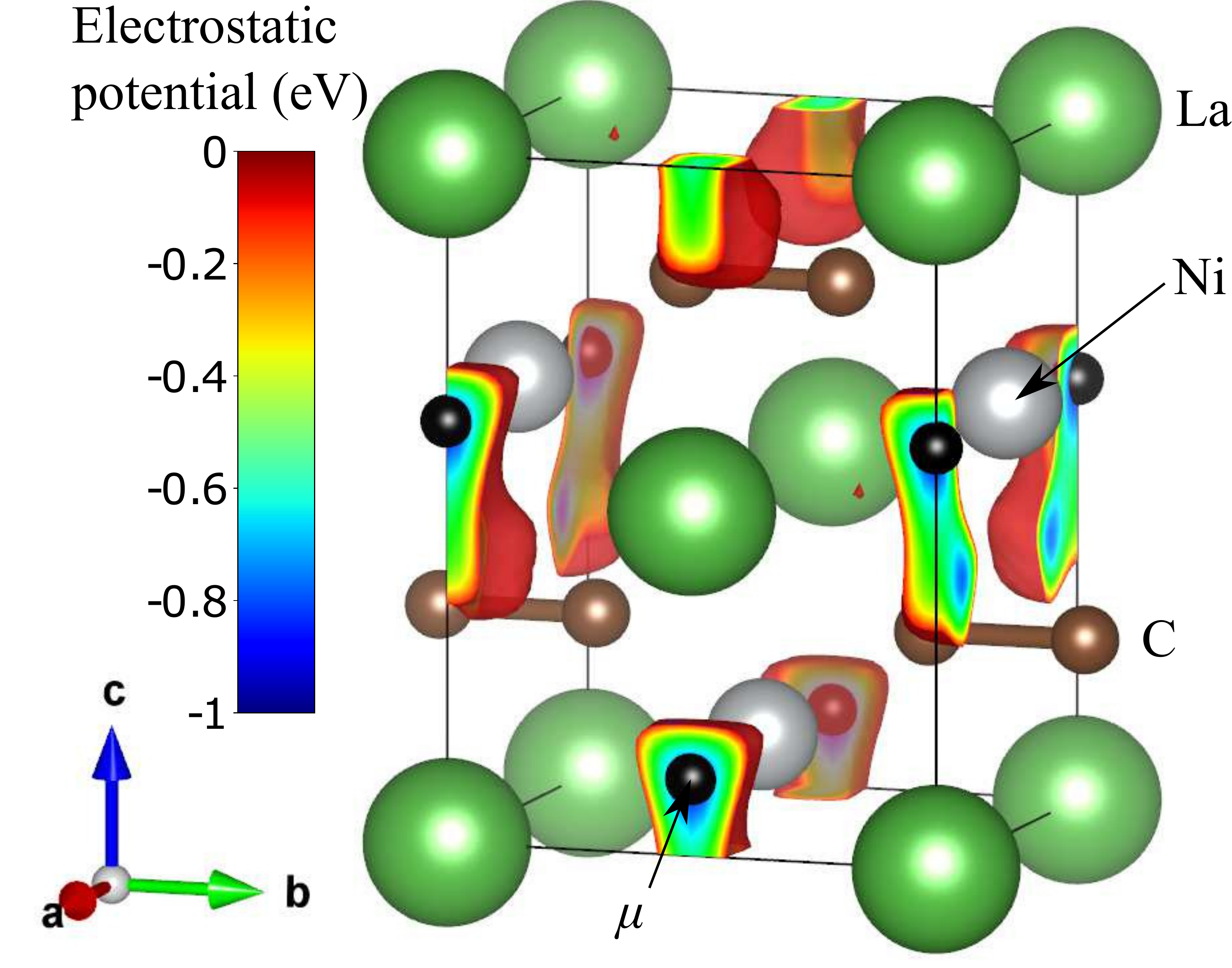}
	\caption{The lowest energy muon stopping site in LaNiC$_2$ obtained from structural relaxations and the electrostatic potential of the host crystal.}
	\label{fig:Fig1}
\end{figure}

For  LaNiC$_2$ we used a plane-wave cutoff energy of 800 eV and a $10 \times 8 \times 6$ Monkhorst-Pack grid \cite{mpgrid1976} for Brillouin zone integration, resulting in total energies that converge to 0.001 eV per cell.  Using these parameters, the optimized cell parameters were found to be within 2.5\% of those determined experimentally \cite{PhysRevLett.102.117007}.  The lattice constants were therefore fixed to their experimental values for subsequent calculations. We first computed the electrostatic potential for the host crystal, as the minima in the electrostatic potential has previously been shown to be a good estimate for the muon stopping site \cite{hideseek} and show this in Fig. \ref{fig:Fig1}.

Structural relaxations were carried out on a supercell comprising $2 \times 2\times2$ conventional unit cells of LaNiC$_2$; a supecell was used to reduce the interaction of the muon and its periodic images.  Due to the enlarged unit cell, we instead used a  $5 \times 4 \times 3$ Monkhorst-Pack grid \cite{mpgrid1976} for these calculations.
Initial structures comprising a muon and the LaNiC$_2$ supercell were generated by requiring the muon to be at least 0.5~\AA~away from each of the muons in the previously generated structures (including their symmetry equivalent positions) and at least 1.0~\AA~away from any of the atoms in the cell.  This resulted in 22 structures which were subsequently allowed to relax.  

These structural relaxations yielded three crystallographically distinct muon stopping sites, summarized in Table \ref{table1}.  We note that sites 2 and 3 are 1.1~eV and 1.6~eV higher in energy than the lowest energy site and are therefore unlikely to be stable stopping sites.  We therefore propose that there is a single crystallographically distinct muon stopping site in this material (site 1). In site 1, the muon is triangularly-coordinated by three La atoms in the $bc$ plane and sits between two Ni atoms along the $a$ axis, with two equal Ni--$\mu^+$ distances of 1.86~\AA.  The Ni atoms are displaced by around 0.125 \AA~towards the muon.  The La atoms are displaced radially outwards, though by a smaller distances (between 0.02~\AA~and 0.06~\AA).  As seen in Fig. \ref{fig:Fig1}, site 1 corresponds to the minimum of the electrostatic potential in LaNiC$_2$.

\begin{table} 
	\begin{tabular}{lccr}
		\hline
		\hline
		Site no. & Energy (eV) & Fractional coordinates & $\sigma_\mathrm{VV}$ (MHz) \\ 
		\hline
		1 & 0 & (0.00441, 0.49766, 0.11178) & 0.121 \\
		2 & 1.1 & (0.23966, 0.27547, 0.23176) & 0.130 \\
		3 &  1.6 & (0.49212, 0.26244, 0.45431) & 0.116 \\
		\hline
		\hline
	\end{tabular}
	\caption{Each of the crystallographically distinct muon stopping sites obtained from structural relaxations and their energies (relative to the lower energy site).\label{table1}}
\end{table}

The muon site obtained here is distinct from the muon site with fractional coordinates (0.5, 0.5, 0) proposed in Ref. \cite{PhysRevLett.102.117007} on the basis of nuclear dipolar fields. We have calculated the relaxation rates corresponding to the Van Vleck second moments \cite{PhysRevB.20.850} for each of the candidate muon stopping sites and report these in the final column in Table~\ref{table1}. The nuclear relaxation rates for all three sites are very similar and not too different from the value $\sigma=0.08$ MHz obtained experimentally \cite{PhysRevLett.102.117007}. 

\begin{figure}[h]
	\includegraphics[width=\columnwidth]{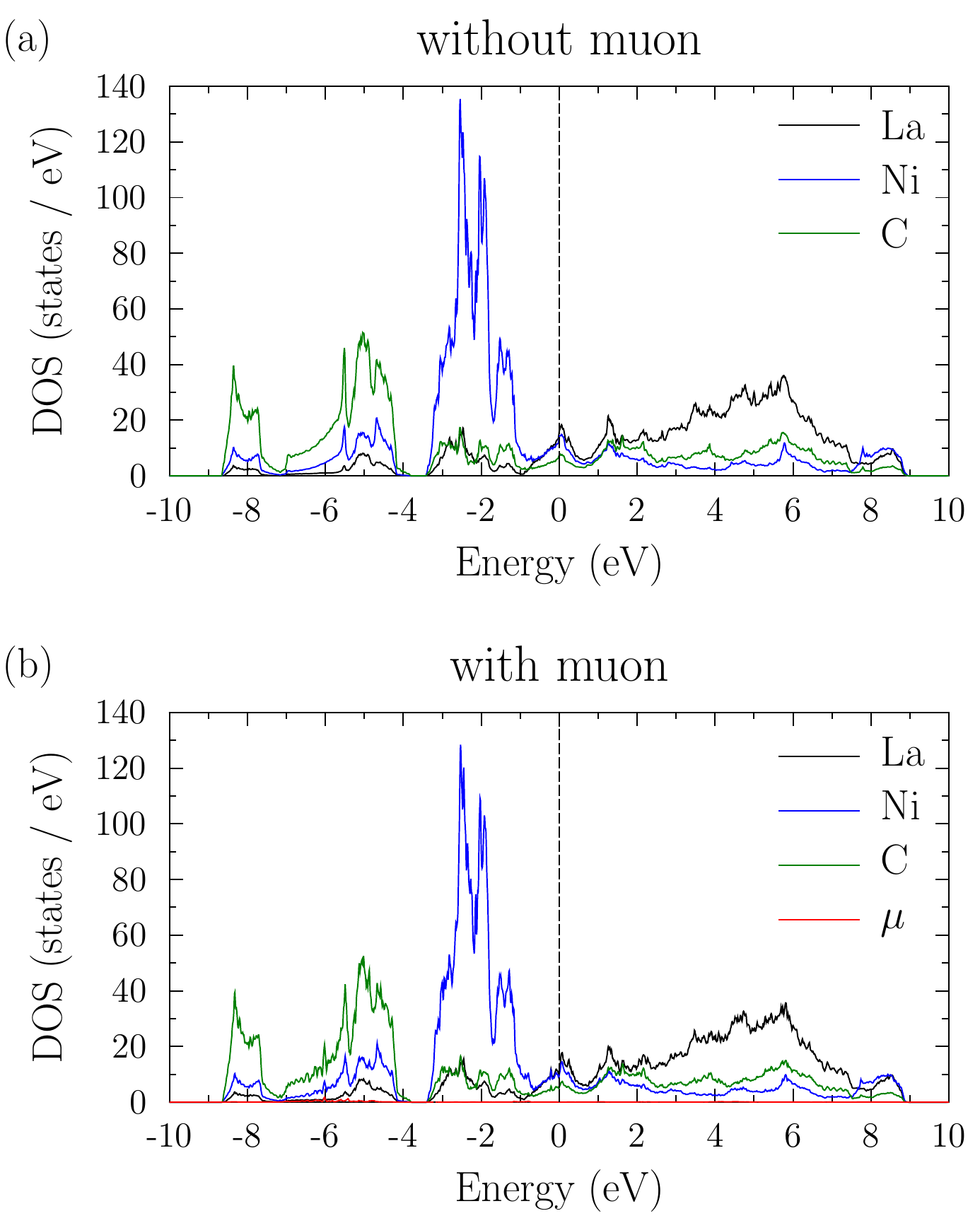}
	\caption{Projected density of states (PDOS) for each atomic species in LaNiC$_2$ (a) without a muon and (b) with a muon. Energies are given relative to the Fermi energy.}
	\label{fig:LaNiC2_dos}
\end{figure}

To investigate the possible effects of the implanted muon of the electronic structure of LaNiC$_2$, we calculated the density of states (DOS) of the system with and without a muon and show this in Fig. \ref{fig:LaNiC2_dos}. A finer $15 \times 12 \times 9$ Monkhorst-Pack grid \cite{mpgrid1976} for  was used for $k$-point sampling when computing the DOS.  The DOS has been projected onto each atomic species in the system and we see that the DOS associated with the muon is 6~eV below the Fermi energy and hence this defect state is not expected to affect the electronic properties of the system.

A possible source of a non-zero local magnetic field would be if the muon induced spin density in its local environment. We therefore carried out a series of spin-polarized calculations. However, these calculations found no appreciable spin density, both for the pristine system and for the system plus implanted muon.  We also investigated the effect of introducing muonium (the bound state of $\mu^+$ and an electron) to the system, as this would introduce an additional ($S=1/2$) electron to the system.  The muon sites for muonium are almost identical to those for $\mu^+$.  We find that the final charge of the muon is the same in both cases, with the additional electron density introduced by muonium being delocalized across the C atoms in the unit cell.  It is interesting to note however that the largest changes in {\it spin} density occur at the positions of the Ni atoms.  This results in a very small increase in spin density at the Ni atoms ($<0.01 \hbar/2$ per Ni atom according to a Mulliken analysis).  However, these effects do not appear to be localised  in the vicinity of the muon and instead solely reflect a change in the number of electrons in the unit cell. 

\subsection{\label{sec:srpras}SrPtAs}
For SrPtAs we used a 16$\times$16$\times$8 Monkhorst-Pack grid~\cite{mpgrid1976} for Brillouin zone sampling and used Marzari-Vanderbilt smearing~\cite{marzari1999} with a width of 0.005 Ry to improve convergence. Structural relaxation of the unit cell results in optimized lattice parameters that are within 1.6\% of those obtained from experiment~\cite{wenski1986}. The DFT-optimized lattice parameters and ionic positions were used for subsequent calculations. The DFT relaxed \spa{} unit cell has space group $P6_3/mmc$ and lattice parameters $a=b=4.31245$ \AA, $c= 9.07819$ \AA. Sr, Pt and As atoms are at fractional coordinates (0.0, 0.0,  0.0), (0.33333, 0.66666,0.25) and (0.33333,0.66666, 0.75), respectively.         

\begin{figure}[!h]
\includegraphics[width=5.0cm]{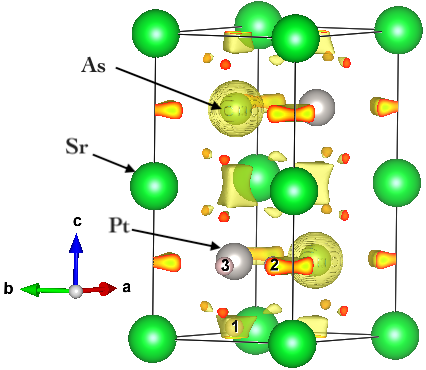}
\centering
\caption[Unitcell ]{\label{fig:musiteelectsr} Isosurface plot of the minimum of the electrostatic potential in the unit cell and the position of the symmetry inequivalent relaxed muon sites labeled 1 to 3. } 
\end{figure}

We first computed the electrostatic potential of the host crystal (shown in Fig.~\ref{fig:musiteelectsr}), as the minima of the electrostatic potential have been shown to provide good approximations to the muon site in a number of cases \cite{PhysRevB.80.094524,De_Renzi_2012,Lamura_2013}. The minimum of the electrostatic potential, together with a grid of positions sampled 1 \AA{} away from the host lattice atoms, form the initial guesses for the muon positions and results in 9 inital muon positions once the crystallographic symmetry is taken into account. Structural relaxations were carried out on a 3$\times$3$\times$2  supercell (108 host atoms and 1 muon)  sampled using a 4$\times$4$\times$4 Monkhorst-Pack grid of $k$-points. Calculations on $\mu^+$ and muonium yielded similar sites, and we therefore present only those sites calculated for $\mu^+$.

\begin{table}
\centering
\caption[muonsite SrptAs]{\label{tab:musrpt} Fractional coordinates of the three  symmetry-inequivalent muon stopping sites in SrPtAs and their energies relative to the lowest energy site. } 
 \begin{tabular}{ l c r }
      \hline
      \hline
      Site no.& Position & Energy (eV)\\ 
       \hline
    1                              & (0.33325,   0.66215,   0.03372)   & 0.00 \\ 
    2                              & (0.08752,   0.20680,   0.24999)   & 0.29 \\ 
    3                             &  (0.03233,   0.51558,   0.24452)  & 0.57 \\  
    \hline
    \hline
    \end{tabular}
\end{table}

\begin{figure}[!h]
\includegraphics[width=8cm, height=2.5cm]{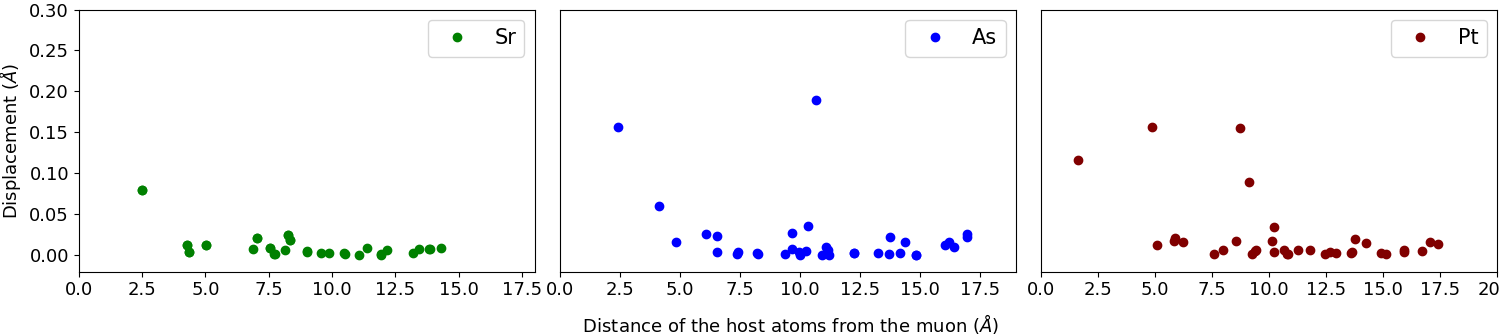}
\caption[Unitcell ]{\label{fig:srptdisappen2}Displacements of the Sr, Pt and As atoms from their equilibrium positions as a function of their distances from muon site 2 in \spa. } 
\end{figure}

Structural relaxations result in three distinct symmetry-inequivalent muon stopping sites.  Their positions and energy differences are reported in Table~\ref{tab:musrpt} and labeled sites 1 to 3. Site 1 has the lowest DFT energy. However, the energy difference of 0.29 eV with site 2 is not sufficiently large to rule it out as a possible stopping site and therefore sites 1 and 2 must both be considered as candidate muon sites. Sites 1 and 2 are situated at minima of the electrostatic potential (see Fig.~\ref{fig:musiteelectsr}), with the global minimum of the electrostatic potential corresponding to the relaxed position of site 1. Both sites 1 and 2 make shorter bond distances with Pt than with Sr and As atoms. In both materials, the  nearest Pt atom to the muon is the most displaced from its equilibrium position with maximum displacement below 0.3 \AA{} (see Fig. 3(b) and Fig.~\ref{fig:srptdisappen2}). 

\begin{figure}[!h]
\includegraphics[width=4.35cm]{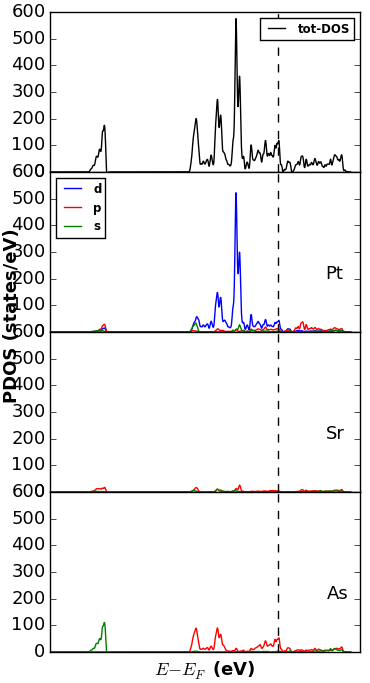}
\centering
\includegraphics[width=4.2cm]{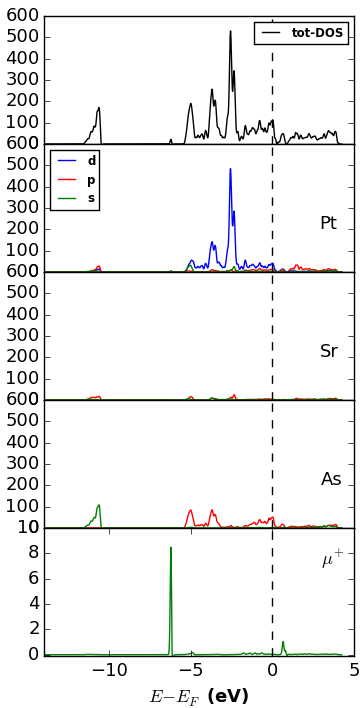}
\centering
\caption[Unitcell ]{\label{fig:srptdosl}Comparison of the projected density of states (PDOS) of the supercells both without (left) and with the muon(right) of site 1  in \spa{}. The Fermi energy has been shifted to 0 eV and is indicated by a dashed line.} 
\end{figure}

The implanted muon does not induce significant changes to the density of state (DOS)  (See  Fig.~\ref{fig:srptdosl}), with the projection of the DOS onto the atomic orbitals showing that the muon states are predominantly far away from the Fermi energy. 
Furthermore, subsequent spin-polarized calculations and L\"{o}wdin charge population~\cite{lowdin1950} analysis do not indicated any muon-induced spin density either at the muon site or for any of the host atoms. This is also the case for muonium; the charge on the muon remains the same within numerical accuracy, while those from the extra electron are fractionally distributed among neighboring As atoms. 

\subsection{\label{sec:zr3ir}\zri{}}
For calculations on the non-centrosymmetric system \zri{}, we used 16$\times$16$\times$8 Monkhorst-Pack grid~\cite{mpgrid1976} for Brillouin zone sampling and used Marzari-Vanderbilt smearing~\cite{marzari1999} with a width of 0.005 Ry to improve convergence. The DFT-optimized lattice parameters were found to be within 1.1\% of those obtained from experiment~\cite{Cenzual1985} and were therefore used for subsequent calculations.  The DFT-relaxed \zri{} unit cell has space group $I\bar{4}2m$ and lattice parameters $a=b=10.82211$ \AA and $c= 5.72236$ \AA. The three Zr and Ir atoms have fractional coordiantess (0.29403, 0.29403, 0.25209), (0.35440, 0.0, 0.5), (0.09572, 0.09572, 0.26267), and (0.29201,0.0, 0.0), respectively. We used the same approach as was used for SrPtAs to generate initial muon positions, obtaining 12 in this case. A 1$\times$1$\times$2 supercell (64 host atoms and 1 muon) and a 4$\times$4$\times$4 Monkhorst-Pack grid was used. The convergence of the supercell size was further confirmed with a 2$\times$2$\times$3 supercell (384 host atoms and 1 muon). Like for SrPtAs, calculations on $\mu^+$ and muonium yielded similar sites, so we present only those sites calculated for $\mu^+$.   

\begin{figure}[!h]
\includegraphics[width=9.0cm]{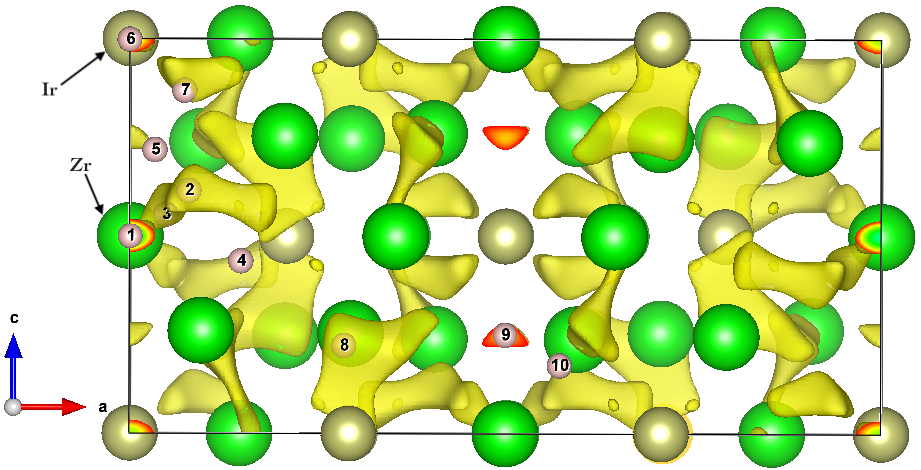}
\centering
\caption[Unitcell ]{\label{fig:musiteelectzr} Isosurface plot of the minimum of the electrostatic potential in \zri{} and the positions of the symmetry-inequivalent relaxed muon stopping sites, labeled 1 to 10.} 
\end{figure}

Stuctural relaxation result in 10 distinct muon sites. The positions and total DFT energy differences of these sites are listed in Table~\ref{tab:efffzri}. These sites are further clustered in 4 groups labeled A, B, C and D by considering the proximity of their positions within the unit cell. As shown in Fig.~\ref{fig:musiteelectzr}, there are a number of distinct minima in the electrostatic potential and most of the calculated muon stopping sites are located at these positions. Despite occupying the global minimum of the electrostatic potential, site 8 does not correspond to the lowest energy relaxed structure, but is instead 0.62 eV higher in energy than the lowest energy site (site 1).

\begin{table}
\centering
\caption[exchnage coupling]{\label{tab:efffzri}Fractional coordinates of the 10  symmetry-inequivalent muon stopping sites, clustered into 4 groups (A,B,C,D) by considering proximity, and their energies (in eV) relative to the lowest energy site. Also shown is the distance of each of the positions relative to lowest energy site within each of clusters.} 
 \begin{tabular}{ lcccr}
 \hline
 \hline
    Cluster  & no.& Position & Energy & Distance (\AA) \\ 
       \hline
  A &  1                               & (0.0013, 0.0001,   0.5001)   & 0.0         & 0.0  \\ 
    & 2                                  & (0.0787, 0.0787,   0.6156)    & 0.01     &  1.4  \\ 
    &  3                                 &  (0.0480, 0.1733,   0.5548)   & 0.14     &   2.0\\  
    & 4                                  & (0.1478, 0.2371, 0.4371)    & 0.21     &   3.0  \\ 
    &  5                                    &(0.0332, 0.2198,   0.7186)  &  0.67   &  2.7     \\   
    \hline
 B& 6                                   & (0.0003, 0.0003,   0.9984)   & 0.35      &   0.0 \\ 
   &   7                                  &  (0.0731, 0.0731,   0.8694)  & 0.43      &  1.3   \\  
  \hline
   C&  8                                 & (0.2853, 0.1063,   0.2231)     & 0.62   &   0.0   \\ 
   \hline
   D&  9                                   & (0.5001, 0.0007,   0.2501)   & 0.64    & 0.0       \\   
     &   10                                 &  (0.5725, 0.0066,   0.1732)   & 0.83   & 0.9 \\
     \hline
     \hline
    \end{tabular}
\end{table}

\begin{figure}[!h]
\includegraphics[width=8.0cm]{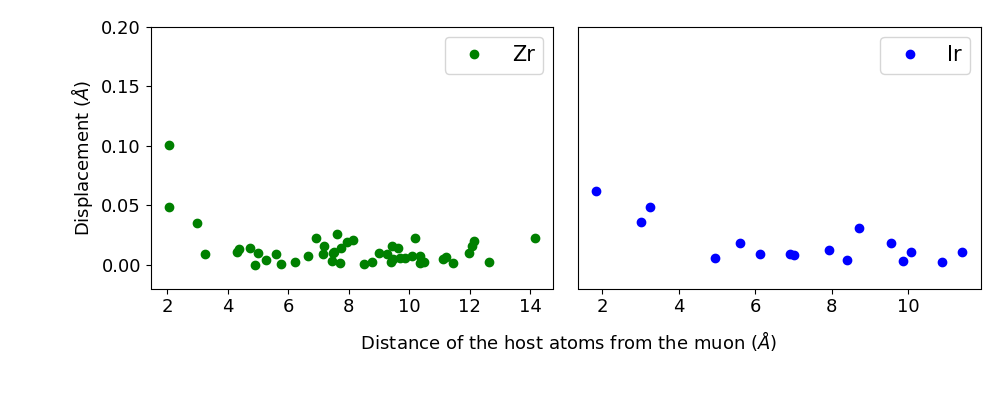}
\centering
\caption[Unitcell ]{\label{fig:displzrsite10}Displacements  of the Zr and Ir atoms from their equilibrium positions as a function of their distances muon site 10 in the 1$\times$1$\times$2 supercell of \zri{} } 
\end{figure}

\begin{figure}[!h]
\includegraphics[width=4.35cm]{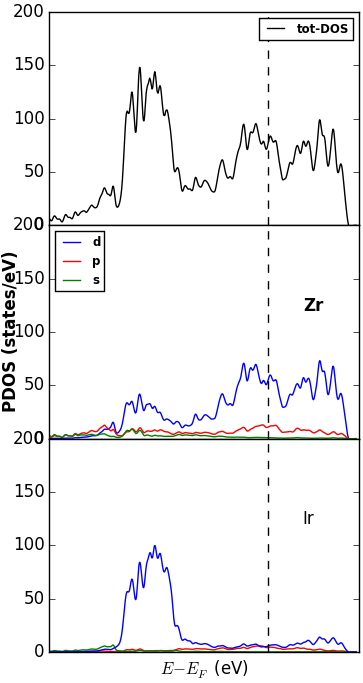}
\centering
\includegraphics[width=4.2cm]{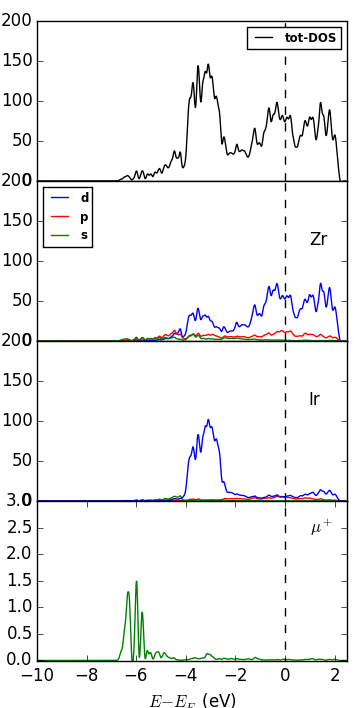}
\centering
\caption[Unitcell ]{\label{fig:zrdos}Comparison of the projected density of states (PDOS) of the supercells both without (left) and with the muon (right) at site 1  in \zri{}. The Fermi energy has been shifted to 0 eV and is indicated by a dashed line. } 
\end{figure}

It is not straightforward to determine which of the site(s) is the stopping position for the muon by considering their energy differences and we therefore instead analyze the effects of the muon by considering a representative site in each of the clusters. The implantation of the muon does not lead to significant distortion of the host Zr and Ir atoms from their equilibrium position, as the maximum displacement remains below 0.1 \AA{} (see  Fig.~3(c) in the main text and Fig.~\ref{fig:displzrsite10}). Furthermore, the muon does not distort the density of states in this system, as the muon states lie far away from the Fermi level (see Fig.~\ref{fig:zrdos}).

\subsection{Re$_6$Zr}
Re$_6$Zr crystallizes in the noncentrosymmetric $\alpha$-Mn structure with cubic space group
$I\bar{4}3m$. The unit cell has 58 atoms that occupy four distinct
crystallographic sites. The stoichiometric composition for the Re-Zr system is Re$_{24}$Zr$_5$, while other compositions have mixed occupancies at these sites \cite{PhysRevB.94.214513}. This makes treating these systems using DFT difficult. The approach we have taken is as follows. Starting with the Re$_{24}$Zr$_5$ structure, we change the lattice parameter to $a=9.714$~\AA~appropriate for Re$_6$Zr \cite{PhysRevB.94.214513}. We then replace the Zr atom at the origin with Re, which has the effect of changing the symmetry of the unit cell from $I\bar{4}3m$ to $P\bar{4}3m$. This gives a composition Re$_{49}$Zr$_9$ and therefore of Re-to-Zr ratio of 5.4:1. Obtaining a Re-to-Zr ratio of exactly 6:1 would require a minimum of 7 unit cells and is therefore computationally prohibitive. Alongside this Re-deficient composition (when compared to Re$_6$Zr) we can also study the effect of being slightly too Re-rich by further substituting the Zr atom at the body-centre of the unit cell with Re. This gives a composition Re$_{25}$Zr$_{4}$ and therefore a Re:Zr ratio of 6.25:1. This additional substitution restores the symmetry to the $I\bar{4}3m$ space group symmetry possessed by Re$_{24}$Zr$_{5}$. The structures corresponding to each of these compositions were allowed to relax, while keeping the input cell fixed. The main difference in between the relaxed structure is that in Re$_{49}$Zr$_9$, that additional Zr atom at the centre of the cell (which is an Re atom for  Re$_{50}$Zr$_8$) repels other nearby Zr atoms away from it. This leads to changes in the precise details of the coordination geometry for muons close to these Zr atoms. Input structures were generated from a single conventional cell of each of these compositions by requiring the muon to be at least 0.5 \AA~away from each of the muons in the previously generated structures (including their symmetry equivalent positions) and at least 1.0~\AA~away from any of the atoms in the cell, which resulted in 43 initial muon positions. For both compositions, we used a plane-wave cutoff energy of 1100 eV and a $4 \times 4 \times 4$ Monkhorst-Pack grid \cite{mpgrid1976} for Brillouin zone integration, resulting in total energies that converge to 0.02 eV per cell.

\begin{table} 
	\begin{tabular}{lcccr}
		\hline
		\hline
		 Site no. & site & nearest & Energy & $\sigma_\mathrm{VV}$  \\ 
		 & geometry & neighbours & (eV) &(MHz) \\
		\hline
		\noalign{\vskip 2mm}
		\multicolumn{5}{c}{Re$_{49}$Zr$_{9}$} \\
		\hline
		A1 & tetrahedral & 3 Re, 1 Zr & 0 & 0.339 \\
		A2 & octahedral & 6 Re & 0.07 & 0.341 \\
		A3 & tetrahedral & 2 Re, 2 Zr & 0.16 & 0.311 \\
		A4 & tetrahedral & 4 Re & 0.38 & 0.386 \\
		\noalign{\vskip 2mm}
			\multicolumn{5}{c}{Re$_{50}$Zr$_{8}$} \\
		\hline
		B1 & tetrahedral & 3 Re, 1 Zr & 0 & 0.336 \\
		B2 & octahedral & 6 Re & 0.07 & 0.338 \\
		B4 & tetrahedral & 4 Re & 0.35 & 0.379 \\
		\hline
		\hline
	\end{tabular}
	\caption{The crystallographically distinct muon stopping sites in each of the compositions approximating Re$_6$Zr obtained from structural relaxations and their energies (relative to the lowest energy site).\label{table_re6zr}}
\end{table}

\begin{figure}[h]
	\includegraphics[width=\columnwidth]{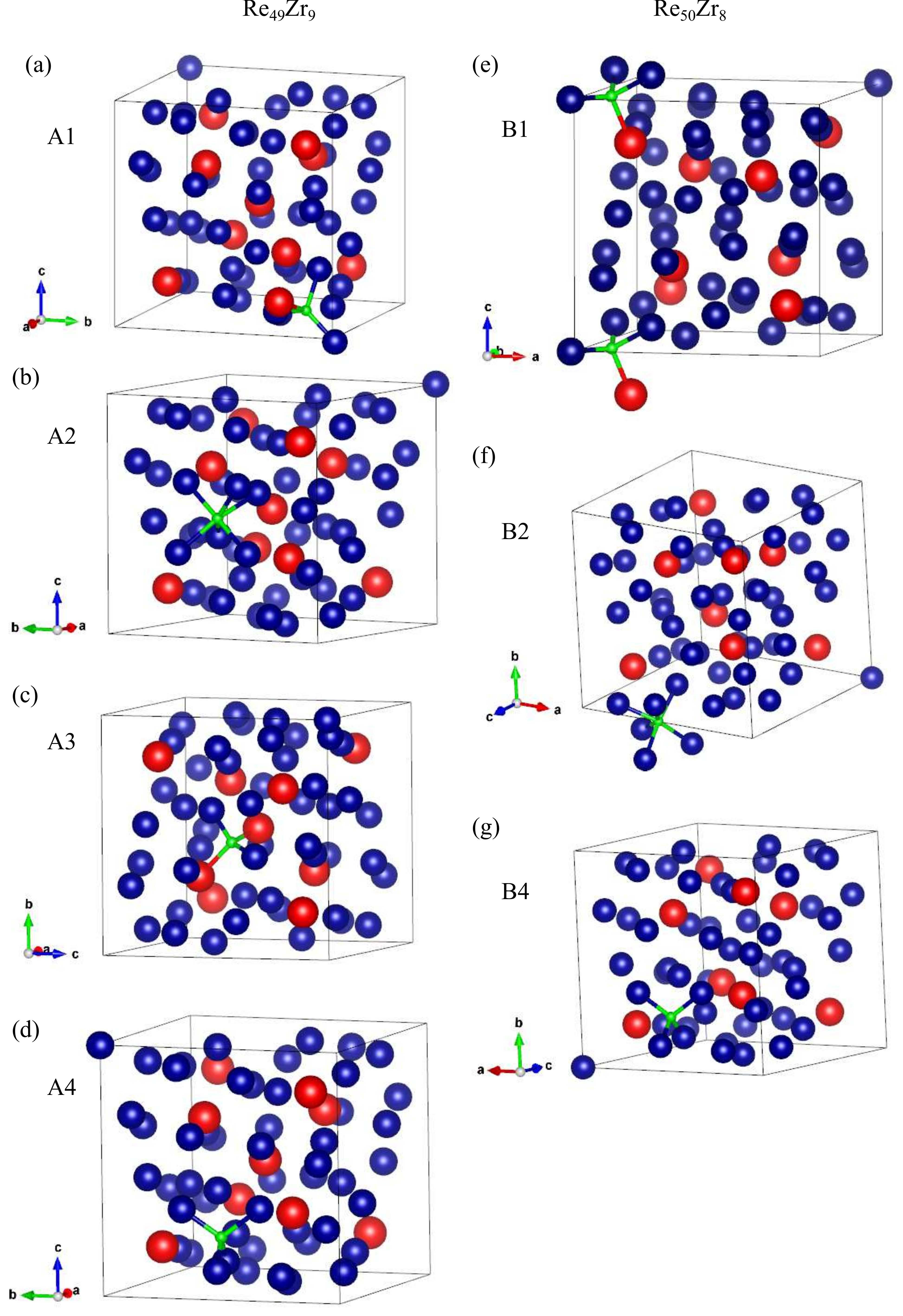}
	\caption{Muon sites in Re$_6$Zr, obtained by considering the close approximations Re$_{49}$Zr$_9$ (left-hand column) and Re$_{50}$Zr$_8$ (right-hand column). Re atoms are blue, Zr atoms are red.}
	\label{fig:re6zr_sites}
\end{figure}

We obtain a large number of crystallographically distinct muon stopping sites after relaxing the initial structures and we summarise the distinct coordination geometries of the muon that we find in Table~\ref{table_re6zr}. The relaxed geometry for each site is shown in Fig.~\ref{fig:re6zr_sites}.
We see that the muon sites in Re$_{49}$Zr$_{9}$ and Re$_{50}$Zr$_{8}$ are almost identical, with the main exception being that a site analogous to A3 in not found in the latter composition. This is due to the fact that this coordination geometry is no longer possible after replacing the Zr atom at the centre of the unit cell with Re. Note that the coordination tetrahedra of the muon in tetahedral sites are not regular (this is true even before the addition of the muon) and this affects the symmetry of the muon site and the muon-induced displacements. For example, for site A1 the muon sits closer to one of the Re atoms in the tetrahedron, which is repelled by 0.16 \AA{}, compared to displacements of around 0.05 \AA{} for the other atoms in the coordination tetrahron (for site B1, the maximum Re displacement is slightly larger at 0.19 \AA{}). In sites where the sits in between four Re atoms (A4 and B4), the local environment of the muon is close to that of a regular tetrahedron; the coordinating Re atoms are repelled 0.08--0.10 \AA{} away from the muon. For the octahedrally coordinated sites, the displacements are slightly smaller, between 0.02 and 0.06 \AA\ for A2 and 0.02--0.07 \AA{} for B2. For site A3 (for which no analogous site was found in Re$_{50}$Zr$_8$), the coordination tetrahedron of the muon is highly irregular, with $\mu^+$--Re distances of 1.80 \AA{} and $\mu^+$--Zr distances of 1.96 \AA{} and 1.96 \AA{}. The muon-induced displacements for this site have similar magnitudes to those found for other tetrahedral sites, with the displacement of the nearest Zr atom of 0.10 \AA{} being the largest Zr displacement found in our calculations.  For all sites, the calculated relaxation rates are somewhat higher that the value $\sigma=0.255$ MHz obtained experimentally \cite{PhysRevLett.121.257002}.

\begin{figure*}[htbp]
	\includegraphics[width=0.8\textwidth]{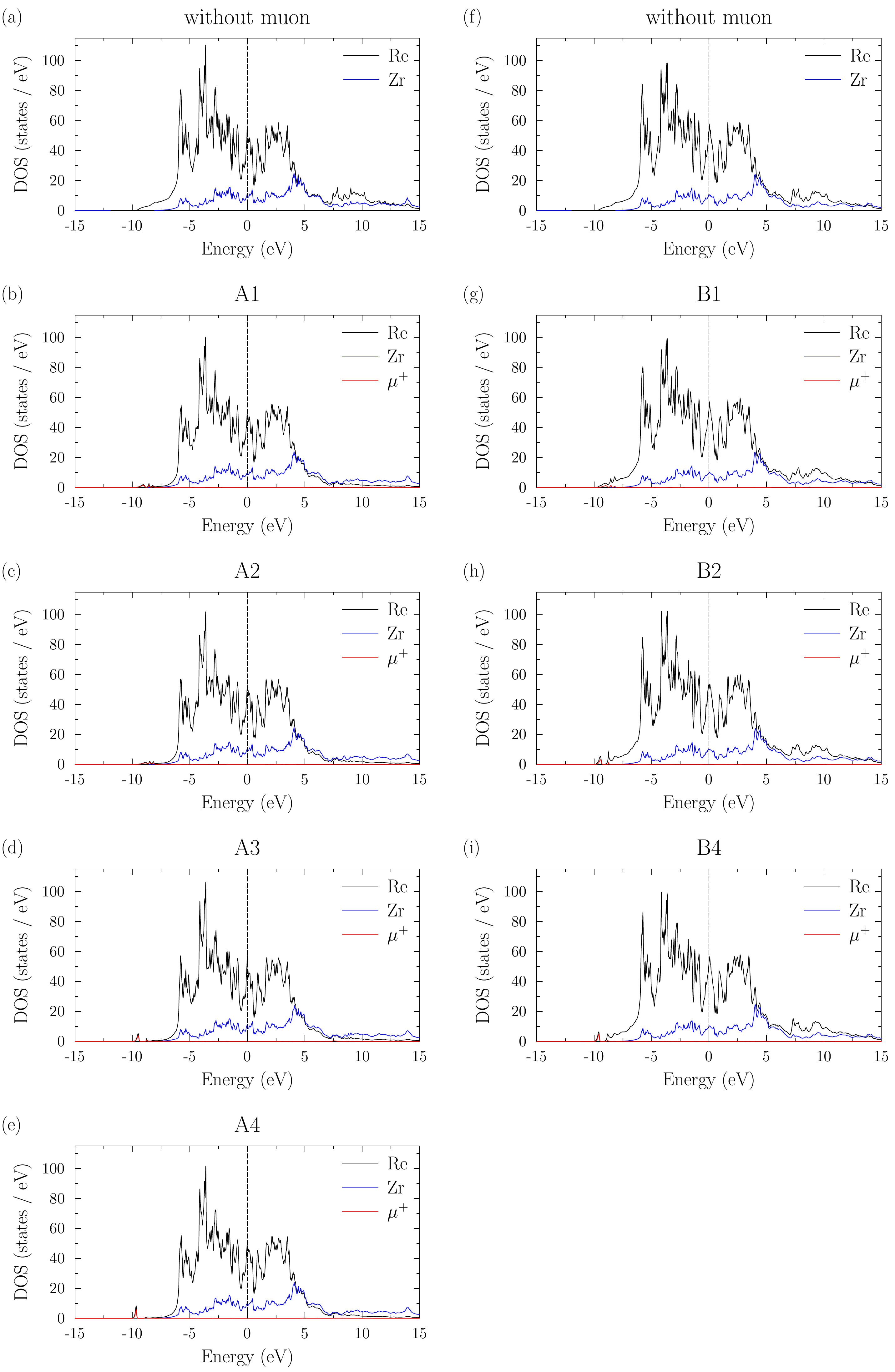}
	\caption{PDOS for each of the muon sites in Re$_{49}$Zr$_{9}$ (left-hand column) and Re$_{50}$Zr$_{8}$ (right-hand column).}
	\label{fig:re6zr_dos}
\end{figure*}

To investigate the effect of the implanted muon on the electronic structure, we have calculated the DOS, with and without the muon, for each of the compositions. The DOS for Re$_{49}$Zr$_{9}$ and Re$_{50}$Zr$_{8}$, shown in Fig.~\ref{fig:re6zr_dos}(a) and Fig.~\ref{fig:re6zr_dos}(h), respectively, are almost identical. We do not observe any significant changes in the DOS close to the Fermi energy. The significant muon DOS all lie at least 8 eV below the Fermi energy for all sites, where we see some hybridization between the muon and Re PDOS. 

\subsection{Niobium}
Niobium crystallizes in the body-centered cubic (bcc) structure with $a=3.30$~\AA. We used a plane-wave cutoff energy of 900 eV and a $18 \times 18 \times 18$ Monkhorst-Pack grid \cite{mpgrid1976} for Brillouin zone integration, resulting in total energies that converge to 1 meV per cell.  The unit cell was allowed to relax and we obtain an optimized lattice parameters $a= 3.31$~\AA, which are within 0.3\% of the experimental values. We used the DFT-optimized lattice parameters and ionic positions in all subsequent calculations.

Structural relaxations were carried out on a supercell comprising $3 \times 3\times3$ conventional unit cells of Nb to reduce the unphysical interaction of the muon and its periodic images.  Due to the enlarged unit cell, we used a  $6 \times 6 \times 6$ Monkhorst-Pack grid \cite{mpgrid1976} for these calculations.
Initial structures comprising a muon and the Nb supercell were generated by requiring the muon to be at least 0.25~\AA~away from each of the muons in the other structures generated (including their symmetry equivalent positions) and at least 1.0~\AA~away from any of the atoms in the cell.  This resulted in 11 structures which were subsequently allowed to relax.  

\begin{figure}[h]
	\includegraphics[width=\columnwidth]{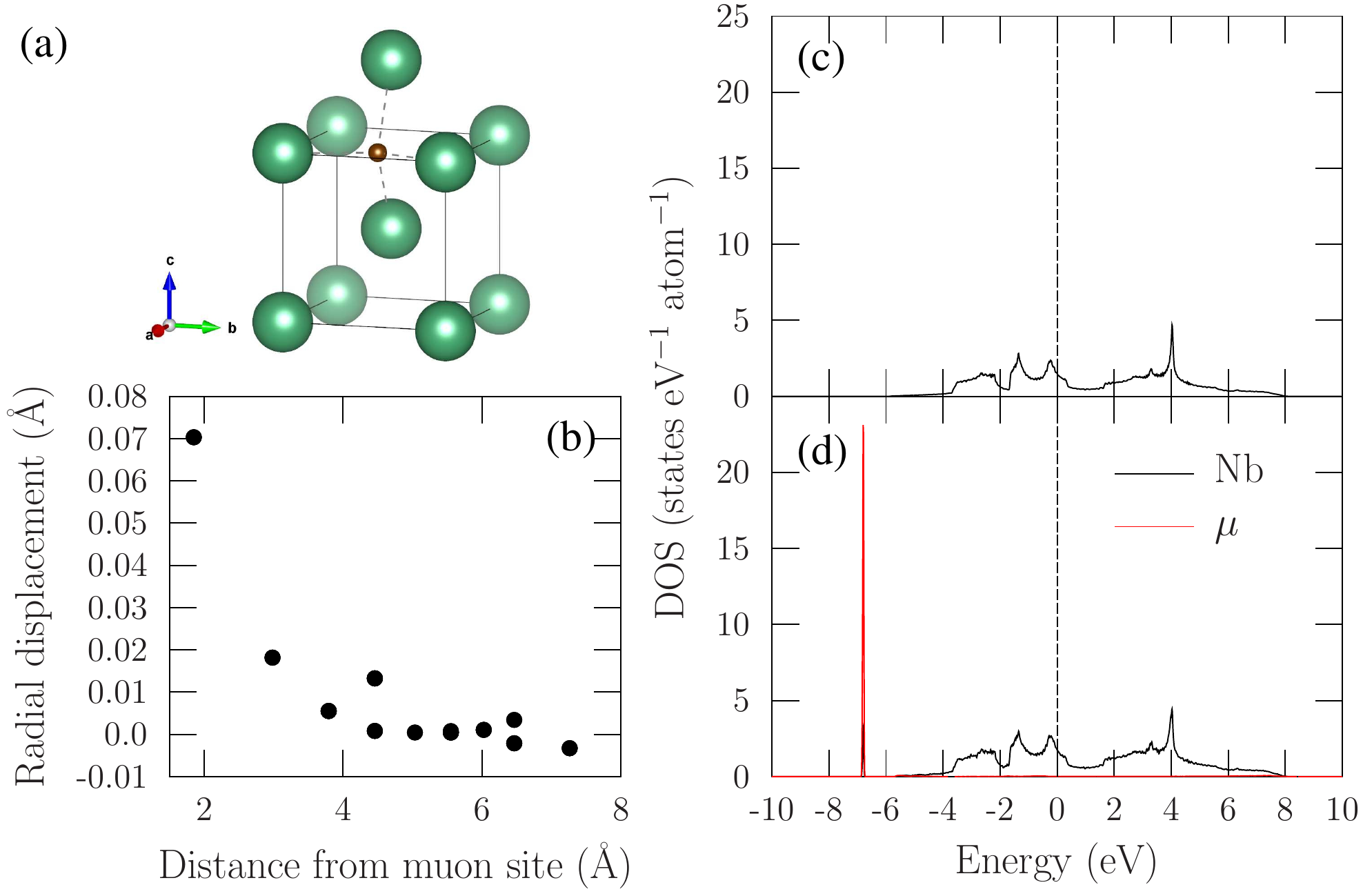}
	\caption{The lowest energy muon site in niobium. (a) The muon occupies a tetrahedral interstitial site of the bcc structure. (b) Radial displacements of the Nb atoms as a function of their distances from the muon site. Projected density of states (PDOS) for the nearest-neighbor Nb atoms (c) without a muon and (d) with a muon.}
	\label{fig:nb_fig}
\end{figure}

These structural relaxations yield a single crystallographically distinct muon stopping sites, shown in Fig.~\ref{fig:nb_fig}(a). The muon occupies a tetrahedral interstitial site of the bcc structure of Nb, with Nb--$\mu^+$ distances of 1.92~\AA. The displacements due to this site are small, with Nb atoms in the coordination octahedron each being repelled by around 0.07~\AA~away from the muon, and with these displacements rapidly decreasing with distance from the muon site, as shown in Fig.~\ref{fig:nb_fig}(b). We note that it is possible to stabilize a muon in an octrahedral interstice, but that this site wasn't obtained from our random search and is 0.28 eV higher in energy than the tetrahedral site.

To investigate the possible effects of the implanted muon on the electronic structure of the system, we computed the density of states (DOS) with and without the muon, for both crystallographically distinct muon sites. We used a finer $18 \times 18 \times 18$ Monkhorst-Pack grid \cite{mpgrid1976} for $k$-point sampling in these calculations. We show the projected density of states (PDOS) for each of the species in the system, without and with an implanted muon, in Figs.~\ref{fig:nb_fig}(c) and ~\ref{fig:nb_fig}(d) respectively. There is some hybridization between this state and the states belonging to the nearest neighbor Nb atoms. There are no significant changes to the DOS in the vicinity of the Fermi energy. The muon density of states take the form of an extremely sharp peak lying around 7 eV below the Fermi energy, reproducing the same muon condition encountered in the superconductors that are found to exhibit TRSB. 

\section{Local moment formation}
	With  each of the systems under study being metals, we can assess
whether the implanted muon acquires an induced moment due to its
interaction with the conduction electrons of the host. The phenomenon
of magnetic impurities is often studied using the Anderson model in
which the energy levels of the impurity (typically one with localized $d$- or $f$-electrons) are broadened into resonances by hybridization with $s$-electrons in the solvent metal. This gives rise to a spectral density function with a Lorentzian form \cite{ziman_1972, coleman_2015}
\begin{equation}
	\rho(E)=\frac{2l+1}{\pi}\frac{W_{l}}{(E-E_\mu)^2+W_{l}^2},
\end{equation}
where $E_\mu$ corresponds to the location of the peak in the muon
density of states. The width $W_{l}$ of the Lorentzian increases with the
strength of the interaction between $s$-electrons and the impurity
states with angular momentum $l$.

The condition for a permanent magnetic moment on the impurity ion is given by \cite{ziman_1972}
\begin{equation}\label{moment_criteron}
	U\rho(E_\mathrm{F})>1,
\end{equation} 
where $U$ is the Coulomb repulsion. In dilute alloys of transition and
rare earth elements, the broadening of the $d$-state often leads to a
reduction in the amplitude of the spectral density that means
$\rho(E_\mathrm{F})$ is too small to satisfy
Eq.~\eqref{moment_criteron}. However, in our case, the muon state is
sufficiently far below the Fermi energy that a significant degree of
broadening is required in order for any appreciable spectral weight to
occur at the Fermi energy. 
We can evaluate the most favorable case for moment formation in which $W_{l}$ takes a value that maximizes the spectral weight of the muon state at the Fermi energy.    

Maximizing $\rho(E_\mathrm{F})$ with respect to $W_{l}$, we find that the largest possible $\rho(E_\mathrm{F})$ is achieved when $W_{l}^2=(E_\mathrm{F}-E_\mu)^2$ for which we obtain
\begin{equation}
	\rho(E_\mathrm{F})=\frac{2l+1}{2\pi}\frac{1}{|E_\mathrm{F}-E_\mu|}
\end{equation}
and hence for a local moment to form we require
\begin{equation}\label{U_condition}
	U>\frac{2\pi}{2l+1}|E_\mathrm{F}-E_\mu|.
\end{equation}

We first consider \SRO. The Coulomb repulsion $U$ can be estimated by
considering the exchange energy for a uniform electron gas
\cite{ziman_1972}, which has a magnitude of 0.916/$r_s$~Ry per
electron, where $r_s$ is the radius of a sphere containing one
electron on average, measured in terms of the Bohr radius. We
approximate the localized state as a ball with a radius equal to half
the distance between the muon and the nearest-neighbor O atom,
$d_{\mu-\mathrm{O}}=0.973$~\AA. The charging energy for occupation of such
a state by two electrons is approximately 30~eV. Using the fact that
the peak in muon DOS is around 8~eV below the Fermi energy,
Eq.~\ref{U_condition} tells us that for $l=0$ local moment formation
is only possible for $U > 50$~eV, which is significantly larger than
our estimate for the Coulomb repulsion energy. The condition for local moment formation could be satisfied by our value of $U$ at higher values of $l$. However, first-principles calculations on atoms in a uniform electron gas show that the Friedel sum is dominated by scattering from the $l=0$ channel for light elements such as the $Z=1$ muon \cite{PhysRevB.27.6121}. Furthermore, $s$-wave scattering dominates the Kondo theory description of muon diffusion in metals \cite{PhysRevB.37.4425,PhysRevB.43.3284}.  It is therefore highly unlikely that a local moment can form on the muon due to its resonance with the conduction elements. For Re, the peak in muon density of states is further below the Fermi energy, leading to a smaller $\rho(E_\mathrm{F})$, and the nearest-neighbor ions are further away, resulting in a smaller charging energy following the approach used for \SRO{}. Local moment formation is therefore even less likely for Re. We note that the same considerations would apply to Nb, for which the peak in the muon DOS lies 7 eV below the Fermi energy, and so local moment formation is similarly unlikely.  Since Nb does not show TRSB and does not exhibit additional relaxation below $T_{\rm c}$ in zero-field $\mu$SR experiments \cite{gygax}, we conclude that the TRSB signal in $\mu$SR experiments in these other superconductors is not connected with the defect level induced by the muon itself.

A further point to consider is that even in cases where local moments form, it is possible that they are screened out by the host metal Fermi sea. In the symmetric Anderson model the Kondo temperature is given by \cite{coleman_2015}
\begin{equation}
	k_\mathrm{B}T_\mathrm{K}=\sqrt{\frac{2U
			W_{l}}{\pi}}\exp\left(-\frac{\pi U}{8 W_{l}}\right).
\end{equation}
At temperatures below $T_\mathrm{K}$, any local moment on the muon
will be screened by the conduction electrons. Strong hybridization
between the impurity states and the conduction electrons gives rise to
a large $W_{l}$ that will enhance $T_\mathrm{K}$. Indeed, our optimal
width of $W_{l}=8$~eV for \SRO{} together with $U=30$~eV results in an
enormous Kondo temperature $T_\mathrm{K}=3 \times 10^5$~K. This shows
that even if local moment formation were possible, the values of
$W_{l}$ required to achieve a significant muon DOS at the Fermi energy
would likely lead to a Kondo temperature that is significantly higher
than the critical temperature for the onset of
superconductivity. Therefore, at the temperatures relevant for
superconductivity, the Fermi sea would screen out the local magnetic
moment, leaving no residual magnetism.

\bibliographystyle{apsrev4-2}
\bibliography{references}